\documentclass[12pt]{article}

\usepackage{amsmath}
\usepackage{amssymb}
\usepackage{hyperref}
\usepackage{graphicx}

\newcommand{\gbl}{[\![\;}
\newcommand{\gbr}{\;]\!]}

\newcommand{\commentstarts}{\begin{centering}
\hspace{-1pt}\vrule\vrule
\begin{minipage}[t]{0.03\linewidth}
\hspace{0.025\linewidth}
\end{minipage}
\begin{minipage}[t]{0.95\linewidth}}
\newcommand{\commentends}{\end{minipage}
\end{centering}
\vspace{7pt}
}

\begin{document}  

\begin{titlepage}

\begin{center}
{\Large\bf $ $ \\ $ $ \\
Vertex operators of ghost number three in Type IIB supergravity
}\\
\bigskip\bigskip\bigskip
{\large Andrei Mikhailov${}^{\dag}$}
\\
\bigskip\bigskip
{\it Instituto de F\'{i}sica Te\'orica, Universidade Estadual Paulista\\
R. Dr. Bento Teobaldo Ferraz 271, 
Bloco II -- Barra Funda\\
CEP:01140-070 -- S\~{a}o Paulo, Brasil\\
}

\vskip 1cm
\end{center}

\begin{abstract}
We study the cohomology of the massless BRST complex of the Type IIB pure 
spinor superstring in flat space. In particular, we find that the cohomology at 
the ghost number three is nontrivial and transforms in the same representation 
of the supersymmetry algebra as the solutions of the linearized classical 
supergravity equations. Modulo some finite dimensional spaces, the ghost 
number three cohomology is the same as the ghost number two cohomology. We
also comment on the difference between the naive and semi-relative cohomology,
and the role of b-ghost.
\end{abstract}

\vfill
{\renewcommand{\arraystretch}{0.8}%
\begin{tabular}{rl}
${}^\dag\!\!\!\!$ 
& 
\footnotesize{on leave from Institute for Theoretical and 
Experimental Physics,}
\\    
&
\footnotesize{ul. Bol. Cheremushkinskaya, 25, 
Moscow 117259, Russia}
\\
\end{tabular}
}

\end{titlepage}

\tableofcontents 

\section{Introduction}\label{sec:Introduction}
Vertex operators are one of the central objects in string theory. They 
represent cohomology classes of the BRST operator. The BRST cohomology 
depends on the chosen background, and in fact describes the tangent space to 
the moduli space of backgrounds at the chosen point. 

In particular, let us look at the pure spinor superstring theory in expansion 
around flat space. The structure of massless BRST cohomology in flat space is 
more or less clear, but it appears that it has never been explicitly spelled 
out in the literature. The present paper is aimed at filling this gap.

For the closed {\em bosonic} string the cohomology was computed in \cite{Astashkevich:1995cv}. We will
here do a similar computation for the pure spinor superstring, but with the 
following difference. It is well known that the physically relevant cohomology 
problem is the so-called semirelative cohomology \cite{Nelson:1988ic}, which is $Q_{\rm BRST}$ acting 
on the vertex operators $V$ satisfying the following condition:
\begin{equation}\label{SemiRelativeCondition}
   (b_0 - \overline{b}_0) V = 0
\end{equation}
This condition was built-in into the computations of  \cite{Astashkevich:1995cv}. In the pure spinor 
superstring, the construction of the $b$-ghost is very subtle. In our paper we 
will compute the ``naive'' cohomology of $Q_{\rm BRST}$, without taking into account 
(\ref{SemiRelativeCondition}). Failure to take into account (\ref{SemiRelativeCondition}) leads to some strange results: 
\begin{enumerate}
\item Nonphysical vertex operators, {\it i.e.} elements of the BRST cohomology 
   which do not correspond to any linearized SUGRA solutions
\item Absence of the dilaton zero mode
\item Nontrivial cohomology at the ghost number three
\end{enumerate}
Problems 1 and 2 are removed if we require the existence of the dilaton 
superfield $\Phi$ (see \cite{Mikhailov:2012id} and the discussion in Section \ref{sec:CommentOnNonPhysicalStates}). To defeat the ghost 
number three cohomology is more difficult. It is dangerous as a potential 
obstacle for continuing an infinitesimal solution to a finite solution 
({\it i.e.} obstructed deformations of the flat spacetime). Such obstructions
would render the theory physically inconsistent. In bosonic string, all 
linearized deformations are unobstructed. One explanation is that the 
{\em semi-relative} cohomology at the ghost number three is zero, and therefore 
there is no obstacle. More precisely, the higher order correction to $V$ are 
controlled by the string field equation \cite{Zwiebach:1992ie,Lian:1992mn}:
\begin{align}\label{StringFieldEquation}
QV = (b_0 - \overline{b}_0)(V^2) + \ldots   
\end{align}
Since the ghost number four cohomology is zero, $V^2$ is in the image of $Q$. In fact,
the pre-image could be chosen to be annihilated by $L_0 - \overline{L}_0$, and this shows that
Eq. (\ref{StringFieldEquation}) can be resolved order by order in the deformation parameter. 

Unfortunately, we do not have such a proof in the pure spinor formalism. It 
follows from the consistency of \cite{Berkovits:2001ue} that there is actually no obstacle in 
extending the infinitesimal deformation to higher orders. Even though the ghost 
number three cohomology is nonzero, the actual obstruction vanishes for physical
states. It would be good to have a transparent proof of this fact using the
language of BRST cohomology and vertex operators. This would probably require
the use of the composite $b$-ghost.

\subsection{Plan of the paper}
In the rest of this introductory section we will review general facts about
the BRST cohomology and its relation to the deformations of the worldsheet
sigma-model. Then in Section \ref{sec:TypeIIBvsMaxwell} we will review the cohomology of the
classical electrodynamics, and explain how to reduce the cohomology of the
Type IIB BRST operator in flat space to the cohomology of electrodynamics. 
The relation will involve the computation of the cohomology of the algebra
of translations with coefficients in the space of solutions of SUSY Maxwell
equations (Section \ref{sec:CohomologyOfClassicalElectrodynamics}) and the tensor produce of two copies of such spaces
(Section \ref{sec:CohomologyOfTensorProduct}). The results on BRST cohomology are summarized in Sections \ref{sec:BRSTCohomology} 
and \ref{sec:ActionOfSUSY}.

\subsection{Classical sigma-model and its deformations}
It was shown in  \cite{Berkovits:2001ue} that classical solutions of the Type IIB supergravity 
are in one-to-one correspondence with two-dimensional sigma-models satisfying
certain axioms. Most importantly, there should be two nilpotent odd symmetries
$Q_L$ and $Q_R$:
\begin{equation}
   Q_L^2 = Q_R^2 = \{Q_L,Q_R\} = 0
\end{equation}
Also, there should be conserved charge known as the ``ghost  number'', with
both $Q_L$ and $Q_R$ having ghost number $+1$.

Suppose that we are given such a sigma-model. A natural question is, how can it 
be deformed? Deformations of the sigma-model are the deformations of the action:
\begin{equation}\label{DeformationOfTheAction}
   S \to S + \varepsilon\int U
\end{equation}
where $U$ is some operator. If $U$ vanishes on-shell, then such deformation is 
trivial, as it can be undone by a field redefinition. Suppose that the
deformation is non-trivial. 

\subsection{From integrated vertex to unintegrated vertex}
The condition that the deformed action still has a pair of nilpotent symmetries 
is equivalent to requiring the existence of $X_L$ and $X_R$ such that on-shell:
\begin{equation}\label{SeparateQLandQR}
   Q_LU \simeq dX_L \;\mbox{ \tt\small and }\; Q_RU \simeq dX_R
\end{equation}
Here $\simeq$ means ``equivalent on-shell'', {\it i.e.} ``equivalent modulo the 
equations of motion''. Explicitly, (\ref{SeparateQLandQR}) implies the existence of infinitesimal
transformations $q_L$ and $q_R$ (vector fields on the field space) such that:
\begin{equation}\label{TotalDerivativeOnShell}
   Q_L U + \varepsilon q_L{\cal L} = d\widetilde{X}_L \;\mbox{ \tt\small and }\;
   Q_R U + \varepsilon q_R{\cal L} = d\widetilde{X}_R
\end{equation}
were ${\cal L}$ is the sigma-model Lagrangian. (The $\widetilde{X}_{L|R}$ of (\ref{TotalDerivativeOnShell}) may be different 
from the $X_{L|R}$ of (\ref{SeparateQLandQR}) because the variation of the Lagrangian is 
proportional to the equations of motion only modulo a total derivative). Then 
$Q_L + \varepsilon q_L$ and $Q_R + \varepsilon q_R$ are both symmetries of the deformed action (\ref{DeformationOfTheAction}). 
Actually they are nilpotent:
\begin{equation}\label{AnticommutatorAutomaticallyZero}
   (Q_L + \varepsilon q_L)^2 =   (Q_R + \varepsilon q_R)^2 = 
\{Q_L + \varepsilon q_L\;,\;Q_R + \varepsilon q_R\} = O(\varepsilon^2)
\end{equation}
This is automatically true because all those anticommutators would be conserved
charges of the ghost number two. In this paper we study vertices which are 
homogeneous polynomials of $x$ and $\theta$. The conserved charges of the ghost number 
two are polynomials of low degree. Therefore if $U$ is of large enough degree 
in $x$ and $\theta$, then the nilpotence condition (\ref{AnticommutatorAutomaticallyZero}) is satisfied.

It is enough to verify (\ref{SeparateQLandQR}) for $Q = Q_L + Q_R$:
\begin{align}\label{Q}
   \exists X\;\mbox{ \tt\small such that }\; Q U = dX
\\    
\mbox{\tt\small where }\; Q = Q_L + Q_R
\end{align}
Conditions (\ref{SeparateQLandQR}) and (\ref{Q}) are equivalent, because $Q_LU$ and $Q_RU$ are 
independent, as both left and right ghost number are conserved. In fact, any
linear combination $\alpha Q_L + \beta Q_R$ with nonzero constant $\alpha$ and $\beta$ can be choosen 
as a BRST operator; all such complexes are quasi-isomorphic to each other. 

Operators $U$ satisfying the condition (\ref{Q}) are called ``integrated vertices''.
Notice that $X$ is a one-form of the ghost number one, and $d(QX) =0$; this
typically\footnote{In this paper we will study vertices which are homogeneous
polynomial of $x$ and $\theta$; some of our results are only valid under the
assumption that the degree of the polynomial is large enough; exceptions may
happen for vertices which do not depend on $x$} implies $QX = dV$,
because there are no conserved charges of the ghost number two. This $V$ is
called the {\em unintegrated vertex} corresponding to the integrated vertex $U$:
\begin{equation}
   (Q_L + Q_R) X = dV
\end{equation}
It is also possible to revert this procedure and go from $V$ back to $U$. This 
involves the assumption about the vanishing of the cohomology at the nonzero 
conformal weight\footnote{Going from the deformation of the action to the cohomology
of $Q_L + Q_R$ requires the absence of local conserved charges with nonzero ghost
number; going back (from $V$ to $U$) requires the vanishing of the cohomology in 
the sector with positive conformal dimension}. Although (to the best of our knowledge) the proof of this 
vanishing theorem has never been  given, we feel that the statement is true. 
Notice that the construction of \cite{Chandia:2013kja} establishes the correspondence between 
integrated and unintegrated vertices independently of this assumption. Although 
(in its current form) it only works in flat space and in $AdS_5\times S^5$, it also 
teaches us something about the generic curved background. For example, it tells 
us that the map $U\mapsto V$ is injective. Indeed, suppose that existed an 
integrated vertex $U$ such that $QU = dX$ and $QX = 0$ ({\it i.e.} nonzero $U$ 
gives $V$). Let us expand such $U$ in Taylor series around a fixed point in the 
curved space-time, and take the leading term. This should give us the flat space
vertex. Since the map $U\mapsto V$ is injective in flat space, the leading term in 
$V$ should also be nonzero. This means that, if $V$ gets killed, then $U$ cannot 
survive either.

\vspace{10pt}

\noindent
In any case, our {\bf working hypothesis} is:
\begin{itemize}
\item at the linearized level the deformations of the action are 
in one-to-one correspondence with the BRST cohomology of $Q = Q_L + Q_R$ at the 
ghost number two
\end{itemize}

\subsection{Ghost number three vertices as obstacles to deformations} 
If $U$ is an integrated vertex operator, then (\ref{DeformationOfTheAction}) defines a deformation of the
sigma-model action to the first order in $\varepsilon$. It is natural to ask, if the 
deformation can be continued to higher orders of $\varepsilon$. An obstacle can, in 
principle, arise already at the order $\varepsilon^2$. Once we deform the action as in 
(\ref{DeformationOfTheAction}), the BRST operator gets deformed:
\begin{equation}
   Q \to Q + \varepsilon q
\end{equation}
Here $q$ is such that:
\begin{equation}
   QU + q{\cal L} = dX
\end{equation}
where ${\cal L}$ is the sigma-model Lagrangian (the existence of such  $q$ follows 
from the fact that $QU$ is a total derivative on-shell, this is in the 
definition of an integrated vertex operator). Let us consider the following
expression: $Q(qX - I_{q^2})$ where $I_{q^2}$ is the Hamiltonian generating 
$q^2$:
\begin{equation}
   q^2{\cal L} = d I_{q^2} \mbox{ \tt\small on-shell }
\end{equation}
It was proven in \cite{Bedoya:2010qz} that exists a ghost-number-three operator $W$ such that:
\begin{align}\label{DefinitionOfObstacle}
   Q (qX - I_{q^2}) \; = &\; dW
\\   
\mbox{\tt\small with } & \;QW = 0
\end{align}
Moreover, the cohomology class of $W$ is the obstacle for extending the
deformation to the order $\varepsilon^2$. The same analysis can be extended to higher orders
in $\varepsilon$.

\paragraph     {Conclusion:} If the BRST cohomology at the ghost number three is 
zero then any infinitesimal deformation can be continued to a finite 
deformation, at least as a power series in $\varepsilon$. However, if the BRST cohomology
at the ghost number three is nonzero, then there is a potential obstacle.

\paragraph     {Comment on the derivation in \cite{Bedoya:2010qz}} 
In \cite{Bedoya:2010qz} we concentrated on the perturbation theory around $AdS_5\times S^5$, while in 
the present paper we work in flat space. Some of the assumptions leading
to Eq. (\ref{DefinitionOfObstacle}) do not work literally in flat space. For example,  conserved
charges with nonzero ghost number (besides the BRST charge) do exist in flat 
space \cite{Mikhailov:2012id}. However, these charges do not depend on $x$. If we restrict ourselves
to the polynomial expressions with large enough degree, then the arguments of
\cite{Bedoya:2010qz} do apply.

\paragraph     {Another way of looking at the obstacle}
Suppose that we have an unintegrated vertex operator $V$ of the ghost number 
two. Suppose that we deform the action as in (\ref{DeformationOfTheAction}) by {\em some} integrated 
operator $\widetilde{U}$ (which is related by the descent procedure to {\em some other} 
integrated vertex $\widetilde{V}$). The BRST operator gets deformed: $Q \mapsto Q + \varepsilon\widetilde{q}$. The
question is, does $V$ survive such a deformation? In other words it is possible
to correct $V\mapsto V + \varepsilon v$ in such a way that $(Q + \varepsilon \widetilde{q})(V + \varepsilon v) = o(\varepsilon^2)$? If the
cohomology at the ghost number three is trivial, then this is always possible.
Otherwise, further analysis is needed: one has to prove that the ghost number
three vertex $\widetilde{q}V$ is $Q$-exact.

\paragraph     {A simpler related phenomenon}
Similar thing happens at the ghost number one. In flat space, there is a 
nontrivial cohomology at the ghost number one, corresponding to the global
symmetries. However, a generic perturbation of the flat space will kill
all this ghost number one cohomology. This is obvious, as generic linearized 
SUGRA solution does not have any global symmetries. What we want to stress, is
the cohomological interpretation of why the ghost number one cohomology gets
killed: the existence of the ghost number two cohomology.

\subsection{Ghost number three cohomology is nonzero}\label{sec:IntroGhostNumberThree}
In this paper we will show that the ghost number three cohomology is nozero. 

The more or less general example of a cohomologically nontrivial ghost number 
three vertex can be obtained as follows. Let us consider a ghost number
two vertex for an exponential linearized solution, for example a Ramond-Ramond 
excitation:
\begin{align}\label{ExponentialV2}
V_2 = e^{(k\cdot x)} \left(
(\theta_L\Gamma^m\lambda_L)(\theta_L\Gamma_m) + 
[\lambda_L\theta_L^{\geq 4}]
\right)_{\alpha}
P^{\alpha\hat{\beta}}   
\left(
(\theta_R\Gamma^m\lambda_R)(\theta_R\Gamma_m) +
[\lambda_R\theta_R^{\geq 4}]
\right)_{\hat{\beta}}
\end{align}
where $P^{\alpha\hat{\beta}}$ is a constant polarization tensor, $\hat{k} P = P\hat{k} = 0$. Suppose that $a_m$ 
is a constant vector such that $(a\cdot k) \neq 0$. Let us consider:

\begin{equation}\label{V3}
   V_3 = \left(a_m(\lambda_L\Gamma^m\theta_L) - a_m(\lambda_R\Gamma^m\theta_R)\right)
V_2
\end{equation}
Notice that $V_3$ is BRST closed. We will prove in Section \ref{sec:DiracDiracSector} that it is not BRST
exact\footnote{Notice that $a_m(\lambda_L\Gamma^m\theta_L) + a_m(\lambda_R\Gamma^m\theta_R) = Q(a\cdot x)$, but the relative sign in (\ref{V3}) is minus. With the
plus sign it would be $Q((a\cdot x) V_2)$}. Also notice that $(\lambda_L\Gamma^m\theta_L) - (\lambda_R\Gamma^m\theta_R)$ is the ghost number one 
unintegrated vertex corresponding to the global conserved charge of 
translations (the momentum of the string). Vertices of the ghost number three
transform in the same representation of the super-Poincare algebra as the 
linearized SUGRA solutions. (In particular, the  obstacle for $V_3$ to be 
BRST-exact is in fact the scalar $(k\cdot a)$, so all the polarization is in $P^{\alpha\hat{\beta}}$.)

The integrated vertex corresponding to (\ref{V3}) can be constructed as follows.
Let $U_2$ be the integrated vertex corresponging to $V_2$. Let $j$ be the conserved
current corresponding to  $(\lambda_L\Gamma^m\theta_L) - (\lambda_R\Gamma^m\theta_R)$:
\begin{equation}
   Qj = d\Big((\lambda_L\Gamma^m\theta_L) - (\lambda_R\Gamma^m\theta_R)\Big)
\end{equation}
Since $U_2$ is an integrated vertex, exists a 1-form $X$ such that
$QU_2 + q{\cal L} = dX$. Let us denote:
\begin{equation}
U_3 = 
\left((\lambda_L\Gamma^m\theta_L) - (\lambda_R\Gamma^m\theta_R)\right) U
- j\wedge X
\end{equation}
We have:
\begin{align}
   QU_3  =\;&
- \left((\lambda_L\Gamma^m\theta_L) - (\lambda_R\Gamma^m\theta_R)\right) dX \;-
\nonumber \\   
\;&
- j\wedge dV_2
- d \left(
   (\lambda_L\Gamma^m\theta_L) - (\lambda_R\Gamma^m\theta_R)
\right)\wedge X \;\simeq 
\nonumber \\    
\simeq \;&\;
 d\Big(j V_2 - 
 \left((\lambda_L\Gamma^m\theta_L) - (\lambda_R\Gamma^m\theta_R)\right) X
\Big)
\end{align}
The next step is:
\begin{align}
   Q\;\Big(jV_2 - 
 \left((\lambda_L\Gamma^m\theta_L) - (\lambda_R\Gamma^m\theta_R)\right) X
\Big) \; = 
\;d \Big( \left((\lambda_L\Gamma^m\theta_L) - (\lambda_R\Gamma^m\theta_R)\right) V_2\Big)
\end{align}
We conclude that $U_3$ is the integrated vertex operator corresponding to $V_3$. It 
is a two-form of the ghost number one.  
 
In this paper we will study {\em polynomial} vertices, {\it i.e.} vertices
depending on $x$ polynomially. The exponential vertices (\ref{ExponentialV2}) and (\ref{V3}) are sums 
of infinitely many polynomial vertices. Indeed, $e^{(k\cdot x)}$ can be decomposed in the 
Taylor series, and the BRST operator preserves the degree of a polynomial (we 
assign degree $1$ to $x$ and degree $1\over 2$ to $\theta$ and $\lambda$). Polynomial vertices are, 
essentially, harmonic polynomials of $x$ dressed with some appropriate 
$\theta$-dependence.

A low degree example of a polynomial vertex of the ghost number three has been
previously constructed in the revised version of \cite{Mikhailov:2012uh}. It is equivalent to the 
linear term in the expansion of $V_3$ in powers of $x$.

\subsection{Cohomology at ghost number four and higher is zero}
We will prove in Section \ref{sec:GhostNumberFour} that the pure spinor cohomology is zero at the 
ghost number four. We have proven in \cite{Mikhailov:2012uh} that the pure spinor cohomology
is zero at the ghost number greater than four. 

This implies that the ghost number three cohomology survives the deformation
from flat space-time to generic curved space-time. (However, in the case of a 
generic curved space-time, there are no ghost number one vertices; therefore 
the construction of Section \ref{sec:IntroGhostNumberThree} does not work.)

\subsection{Argument for vanishing of the obstruction based on symmetry}
Consider an unintegrated vertex operator $V$ and the corresponding deformation 
of the sigma-model. Can we extend it to the second order in the deformation 
parameter? The potential obstacle is the ghost number 3 cohomology class $W$ 
defined in Eq. (\ref{DefinitionOfObstacle}). It is bilinear in $V$:
\begin{equation}\label{WIsGB}
   W = \gbl V,V \gbr
\end{equation}
We will show that $W$ transforms in the linearized supergravity multiplet 
({\it i.e.} in the same representation as $V$, modulo some discrete states). 
The map $V\otimes V \to W$ given by (\ref{WIsGB}) defined by (\ref{WIsGB}) should commute with the 
action of the supersymmetry, in particular with the translations. Moreover, 
one can see that:
\begin{align}
   \mbox{deg}(W) = 2\;\mbox{deg}(V) - 2
\end{align}
(For example, for the linear dilaton background analized in \cite{Mikhailov:2012id}, $V\simeq [\lambda^2\theta^4]$ 
and $q \simeq \left[\lambda\theta^2{\partial\over\partial\theta}\right]$.) This implies that $\gbl V_1,V_2 \gbr$ can only be nonzero if either 
$V_1$ or $V_2$ is a low degree polynomial. 

It should be possible to complete this argument, which would provide a proof of
the vanishing of the obstructions to most of the deformations of the flat space
at the second order (but this proof will not work at higher orders).

\subsection{Plan of the paper}
In Section \ref{sec:TypeIIBvsMaxwell} we explain how to compute the massless BRST cohomology of the 
Type II SUGRA by relating it to the BRST cohomology of the Maxwell theory using
the spectral sequence of a bicomplex. In Sections \ref{sec:CohomologyOfClassicalElectrodynamics},\ref{sec:ZerothCohomology},\ref{sec:CohomologyOfTensorProduct} and \ref{sec:SecondCohomologyTensorProduct} we compute 
the second page of that spectral sequence. In Section \ref{sec:BRSTCohomology} we finally compute
the spectrum of massless states, and in Section \ref{sec:ActionOfSUSY} we study the action of
supersymmetry on the ghost number three vertices.

\noindent
For the first reading, we recommend the following sequence:

\noindent
Section \ref{sec:Introduction} $\longrightarrow$ Section \ref{sec:TypeIIBvsMaxwell} $\longrightarrow$ Section \ref{sec:BRSTCohomology}.

\noindent
Then Sections  \ref{sec:CohomologyOfClassicalElectrodynamics},\ref{sec:ZerothCohomology},\ref{sec:CohomologyOfTensorProduct} and \ref{sec:SecondCohomologyTensorProduct} could be read at the second pass.

\section{Type IIB BRST complex vs Maxwell complex}\label{sec:TypeIIBvsMaxwell}
We will compute the cohomology of the Type IIB BRST complex by relating it to
the super-Maxwell BRST complex.

\subsection{Super-Maxwell BRST complex}
The cohomology of the super-Maxwell BRST complex:
\begin{equation}\label{QSuperMaxwell}
   Q_{\rm SMaxw} = \lambda^{\alpha}\left(
      {\partial\over\partial \theta^{\alpha}} +
      \Gamma^m_{\alpha\beta}\theta^{\beta}{\partial\over\partial x^m}
   \right)
\end{equation}
is only nontrivial at the ghost numbers 0 and 1. 
At the ghost number 0 the cohomology is constants: $V(\theta_L,\theta_R,x) = \mbox{const}$.
At the ghost number 1, the cohomology is in one-to-one correspondence with the
solutions of the free Maxwell equation and the free Dirac equation. The 
vanishing of the cohomology at the ghost numbers two and three is equivalent to 
the following statements: 
\begin{enumerate}
\item For any current $j_m$ such that $\partial_mj_m=0$ always exists the gauge field $F_{mn}$ 
satisfying $\partial_{[k}F_{lm]}=0$ and $\partial_mF_{mn} = j_n$
\item For any antichiral spinor $\psi$ exists a chiral spinor $\phi$ such that $\Gamma^m\partial_m\phi = \psi$
\item For any $\rho$ exists $j_m$ such that $\partial_m j_m = \rho$
\end{enumerate}
\paragraph     {Example:} Let us look at the ghost number two cohomology. The
leading term in the $\theta$-expansion is either $(\theta\Gamma^m\lambda)(\theta\Gamma^n\lambda)(\theta\Gamma_{mn}\psi(x))$ or
$(\theta\Gamma^m\lambda)(\theta\Gamma^n\lambda)(\theta\Gamma_{mnl}\theta)A^l(x)$. Let us for example investigate the first 
possibility. The following expression is in the cohomology of $\lambda^{\alpha}{\partial\over\partial \theta^{\alpha}}$:
\begin{equation}\label{ZeroModeGhostNumberTwo}
   (\theta\Gamma^m\lambda)(\theta\Gamma^n\lambda)(\theta\Gamma_{mn}\psi(x))
\end{equation}
Now let us study the effect of the $\partial\over\partial x$-term in (\ref{QSuperMaxwell}). For (\ref{ZeroModeGhostNumberTwo}) to survive the
action of $(\lambda\Gamma^m\theta){\partial\over\partial x^m}$ we need:
\begin{equation}
(\lambda\Gamma^l\theta){\partial\over\partial x^l}     
(\theta\Gamma^m\lambda)(\theta\Gamma^n\lambda)(\theta\Gamma_{mn}\psi(x)) =
\lambda^{\alpha}{\partial\over\partial\theta^{\alpha}}\mbox{\tt\small (something)}
\end{equation}
The ``something'' on the right hand side always exists, because any expression
of the form $[\lambda^3\theta^4]$ annihilated by $\lambda^{\alpha}{\partial\over\partial\theta^{\alpha}}$  is automatically in the image of
$\lambda^{\alpha}{\partial\over\partial\theta^{\alpha}}$. It remains to investigate the possibility of (\ref{ZeroModeGhostNumberTwo}) being $Q$-exact:
\begin{align}
\;&
(\theta\Gamma^m\lambda)(\theta\Gamma^n\lambda)(\theta\Gamma_{mn}\psi(x))  \; =
\nonumber \\   
=\;& (\lambda\Gamma^l\theta){\partial\over\partial x^l} 
\Big(
   (\theta\Gamma^k\lambda)(\theta\Gamma_k\phi(x)) + 
\mbox{\small\tt (terms of higher orders in $\theta$)}
\Big) \; +
\nonumber \\  
\;& + \; 
\lambda^{\alpha}{\partial\over\partial\theta^{\alpha}}\mbox{\small\tt (something)} 
\end{align}
This is possible iff $\psi(x) = \Gamma^m{\partial\over\partial x^m}\phi(x)$. But for any $\psi(x)$ we can find $\phi(x)$ 
such that $\psi(x) = \Gamma^m{\partial\over\partial x^m}\phi(x)$. This implies that any expression of the type 
(\ref{ZeroModeGhostNumberTwo}) is always  BRST-trivial. The class with the leading term 
$(\theta\Gamma^m\lambda)(\theta\Gamma^n\lambda)(\theta\Gamma_{mnl}\theta)A^l(x)$ is analyzed similarly.

\paragraph     {Conclusion:}
\begin{align}
   H^0(Q_{\rm SMaxw}) \;& = {\bf C}
\label{SMaxw0}\\   
H^1(Q_{\rm SMaxw}) \;& = {\rm Maxwell}\bigoplus {\rm Dirac}
\label{SMaxw1}\\  
H^{>1}(Q_{\rm SMaxw}) \;& = 0
\label{SMaxwHigher}
\end{align}
Here ``${\rm Maxwell}\bigoplus {\rm Dirac}$'' stands for the direct sum of the space of solutions
of the Maxwell equations and the space of solutions of the Dirac equation.

We now want to relate the super-Maxwell complex to the Type IIB SUGRA complex.

\paragraph     {Comment in the revised version}
It is possible to modify the definition of the BRST complex by imposing the 
constraint that the cochains are annihilated by $L_0+ \overline{L}_0$. In this case 
$H^2(Q_{\rm SMaxw})$ is nonzero and in fact isomorphic (perhaps modulo some zero modes)
to $H^1(Q_{\rm SMaxw})$ --- see the recent work \cite{Jusinskas:2015eza} and references there. 
We do not impose any such constraints. Therefore our BRST complex has 
$H^2(Q_{\rm SMaxw})=0$ for open strings. But for closed strings, we still get 
the massless $H^3(Q_{\rm SUGRA})$ nonzero (and isomorphic to $H^2(Q_{\rm SUGRA})$ up to
zero modes).

\subsection{Definition of the doubled complex}
Let us consider the tensor product of two SMaxwell complexes:
\begin{equation}\label{BRSTSMaxwTimesSMaxw}
Q_{{\rm SMaxw}\otimes {\rm SMaxw}} =
   \lambda^{\alpha}_L \left(
      {\partial\over\partial \theta_L^{\alpha}} +
      \Gamma^m_{\alpha\beta}\theta_L^{\beta}{\partial\over\partial x_L^m}
   \right)
+
   \lambda^{\hat{\alpha}}_R \left(
      {\partial\over\partial \theta_R^{\hat{\alpha}}} +
      \Gamma^m_{\hat{\alpha}\hat{\beta}}
      \theta_R^{\hat{\beta}}{\partial\over\partial x_R^m}
   \right)
\end{equation}
The operator $Q_{{\rm SMaxw}\otimes {\rm SMaxw}}$ acts on the space of functions 
$F(\lambda_L,\lambda_R,\theta_L,\theta_R,x_L,x_R)$. We will denote $Q_L$ and $Q_R$ the two terms on the
right hand side of (\ref{BRSTSMaxwTimesSMaxw}). This is the ``doubled'' BRST complex. The difference
with the Type IIB SUGRA BRST complex is the splitting $x = x_L + x_R$.
In the Type IIB BRST complex there is no separation of $x$ into $x_L$ and $x_R$:
\begin{equation}\label{QSUGRA}
Q_{\rm SUGRA} =
   \lambda^{\alpha}_L \left(
      {\partial\over\partial \theta_L^{\alpha}} +
      \Gamma^m_{\alpha\beta}\theta_L^{\beta}{\partial\over\partial x^m}
   \right)
+
   \lambda^{\hat{\alpha}}_R \left(
      {\partial\over\partial \theta_R^{\hat{\alpha}}} +
      \Gamma^m_{\hat{\alpha}\hat{\beta}}
      \theta_R^{\hat{\beta}}{\partial\over\partial x^m}
   \right)
\end{equation}
The difference with (\ref{BRSTSMaxwTimesSMaxw}) is that the left and the right parts have a common 
$x$ instead of separate $x_L$ and $x_R$; the operator $Q_{SUGRA}$ acts on the space of 
functions $F(\lambda_L,\lambda_R,\theta_L,\theta_R,x)$.

The computation of the cohomology of (\ref{BRSTSMaxwTimesSMaxw}) is straightforward, because it is 
just the tensor product of two Maxwell complexes (\ref{QSuperMaxwell}); therefore the cohomology
is:
\begin{equation}
   H^n(Q_{{\rm SMaxw}\otimes {\rm SMaxw}}) = \bigoplus\limits_{p+q = n} 
H^p(Q_{\rm SMaxw})\otimes H^q(Q_{\rm SMaxw})
\end{equation}
where the spaces $H^p(Q_{\rm SMaxw})$ are given by Eqs. (\ref{SMaxw0}), (\ref{SMaxw1}) and (\ref{SMaxwHigher}).

\subsection{Spectral sequence ${\cal E}_r^{p,q}$} 
To compute the cohomology of (\ref{QSUGRA}), we relate it to the cohomology of (\ref{BRSTSMaxwTimesSMaxw}) by 
the following trick. Let us introduce a formal fermionic variable $c^m$ and the
operator:
\begin{equation}
   Q_{\rm Lie} = c^m\left({\partial\over\partial x_L^m} - {\partial\over\partial x_R^m}\right)
\end{equation}
(We call it $Q_{\rm Lie}$ because it can be thought of as the cohomology of the abelian 
Lie algebra of translations.) Let us consider the bicomplex:
\begin{equation}
Q_{\rm tot} = Q_L + Q_R + Q_{\rm Lie}
\end{equation}
Consider two ways of computing the cohomology of $Q_{\rm tot}$. We can either
compute first the cohomology of $Q_{\rm Lie}$, and then consider $Q_L + Q_R$ as a
perturbation. Or, first compute $H(Q_L + Q_R)$ and then act on it by $Q_{\rm Lie}$. This
means that there are two different spectral sequences, both converging to 
$H(Q_{\rm tot})$. 

\paragraph     {First $Q_{\rm Lie}$, then $Q_L + Q_R$:}
Because of the Poincare lemma, the cohomology of $Q_{\rm Lie}$ is only nontrivial in 
the ghost number 0, and is represented by the functions 
$f(\lambda_L,\lambda_R,\theta_L,\theta_R,x_L + x_R)$. Therefore the Type IIB BRST complex is equivalent 
to the cohomology of $Q_L + Q_R$ acting on the cohomology of $Q_{\rm Lie}$:
\begin{equation}
H(Q_{\rm SUGRA}) = H(Q_L + Q_R\;,\;H(Q_{\rm Lie})) = H(Q_{\rm tot})
\end{equation}

\paragraph     {First $Q_L + Q_R$, then $Q_{\rm Lie}$:}
now let us first compute the cohomology of $Q_L + Q_R$, and then consider $Q_{\rm Lie}$ as
a perturbation. The resulting spectral sequence will be denoted ${\cal E}_r^{p,q}$.
It computes the cohomology of the SUGRA BRST complex:
\begin{align}
   {\cal E}_1^{p,q} \;&= H^p\left(
      Q_{\rm Lie}\;,\; 
      \bigoplus\limits_{q_L + q_R = q} H^{q_L}(Q_L)\otimes H^{q_R}(Q_R)
   \right)
\\   
{\cal E}_r^{p,q}\;&\Rightarrow_p {\cal E}_{\infty}^{p,q}
\\   
\bigoplus_{p+q = n}{\cal E}_{\infty}^{p,q} \;& = H^n(Q_{\rm SUGRA})
\end{align}
Therefore, the only nontrivial components are:
\begin{align}
   {\cal E}_1^{p,0} \;& = \Lambda^p {\bf C}^{10}
   \\    
   {\cal E}_1^{p,1} \;& = H^p(Q_{\rm Lie}\;,\; {\rm SMaxw}_L \bigoplus {\rm SMaxw}_R)
   \\   
   {\cal E}_1^{p,2} \;& = H^p(Q_{\rm Lie}\;,\; {\rm SMaxw}_L \otimes {\rm SMaxw}_R)
\end{align}
All other components are zero. The only potentially nonzero differentials are:
\begin{align}
   {\cal E}_1^{p,0}\xrightarrow{d_1} {\cal E}_1^{p+1,0}\;,\;
   {\cal E}_1^{p,1}\xrightarrow{d_1} {\cal E}_1^{p+1,1}\;,\;
   {\cal E}_1^{p,2}\xrightarrow{d_1} {\cal E}_1^{p+1,2}
\\    
{\cal E}_2^{p,2}\xrightarrow{d_2} {\cal E}_2^{p+2,1}\;,\;
{\cal E}_2^{p,1}\xrightarrow{d_2} {\cal E}_2^{p+2,0}
\\   
{\cal E}_3^{p,2}\xrightarrow{d_3} {\cal E}_3^{p+3,0}
\end{align}

\hspace{-25pt}\includegraphics[scale=0.36]{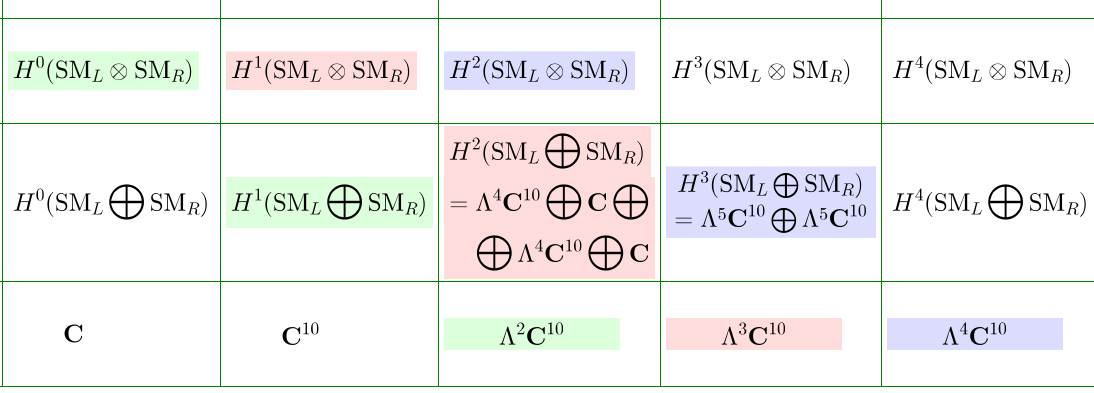} 

\noindent
Therefore, in order to compute the BRST cohomology of SUGRA, we have to: 
\begin{itemize}
\item first compute 
the cohomology of $Q_{\rm Lie}$ with coefficients in spaces of solutions of the 
classical electrodynamics and their tensor products
\item then compute the differentials $d_r$
\end{itemize}
The first step will be elaborated in Sections \ref{sec:CohomologyOfClassicalElectrodynamics} , \ref{sec:ZerothCohomology}
and \ref{sec:CohomologyOfTensorProduct}, and the second in Section \ref{sec:BRSTCohomology}.

\vspace{15pt}
{\large
\noindent The reader may want to {\bf skip to Section \ref{sec:BRSTCohomology}} and return here
later.
}

\section{Cohomology of classical electrodynamics}\label{sec:CohomologyOfClassicalElectrodynamics}
In the previous section we related the cohomology of the SUGRA complex to the
Lie algebra cohomology of the algebra of translations ${\bf R}^{10}$ with coefficients
in the tensor product of solutions of Maxwell and Dirac equations. In order
to compute it, we will first compute the cohomology with coefficients in the 
single space of solutions of Maxwell and Dirac  equations. Then, in the next
section, we will proceed to compute the cohomology with coefficients in the 
tensor product of two such spaces.

\subsection{Cohomology of ${\bf R}^{10}$ with values in solutions of Maxwell equations}\label{sec:CohomologyInMaxwell}
Consider the space of solutions of the vacuum Maxwell equations:
\begin{equation}
   {\partial\over\partial x^m}{\partial\over\partial x^{[m}}A_{n]} = 0
\end{equation}
depending on a parameter $c^m$, a free Grassmann variable. We need to calculate
the cohomology of the operator $c^m{\partial\over\partial x^m}$ acting on this space. 

We will start by computing the cohomology of divergenceless currents.
Consider the space $J$ of one-forms $j_m(x,c)dx^m$ satisfying ${\partial\over\partial x^m}j_m(x,c) = 0$. 
This is a subspace of the space of all 1-forms $\Omega^1$:
\begin{equation}
   0\rightarrow J \xrightarrow{\subset} \Omega^1 \xrightarrow{\delta} \Omega^0 \rightarrow 0
\end{equation}
This gives the long exact sequence of cohomology:
\begin{align}
   0\rightarrow {\bf C}^{10} \rightarrow {\bf C}^{10}\rightarrow {\bf C}
\rightarrow H^1(J)\rightarrow 0\rightarrow 0 \rightarrow H^2(J)
\rightarrow 0\rightarrow\ldots 
\end{align}
We conclude:
\begin{align}
H^0(J) = \;& {\bf C}^d
\\   
   H^1(J) =\;& {\bf C}
\\   
H^{>1}(J) = \;&0
\end{align}
Now we proceed to the cohomology of the Maxwell complex. A solution of the 
Maxwell equation is completely characterized by its curvature. The space of 
solutions is therefore the same as the space of closed 2-forms $F_{mn}dx^m\wedge dx^n$ 
satisfying $\partial^mF_{mn} = 0$. It is included in the following short exact sequence: 
\begin{equation}
   0 \rightarrow F \rightarrow Z^2 \rightarrow J \rightarrow 0
\end{equation}
where $Z^2$ is the space of all closed 2-forms. The corresponding long exact 
sequence reads:
\begin{align}
\longrightarrow \;& 
\Lambda^2{\bf C}^d \longrightarrow \Lambda^2{\bf C}^d 
\longrightarrow {\bf C}^d \longrightarrow
\nonumber \\    
\longrightarrow \;& 
H^1({\rm Maxwell}) 
\longrightarrow H^1(Z^2) \longrightarrow {\bf C} \longrightarrow
\nonumber \\   
\longrightarrow \;& 
H^2({\rm Maxwell}) \longrightarrow H^2(Z^2) \longrightarrow 0 \longrightarrow \ldots
\label{LongExactSequenceForMaxwell}
\end{align}
To calculate the cohomology of $Z^2$ we use:
\begin{equation}
0\longrightarrow Z^1\longrightarrow \Omega^1 \longrightarrow Z^2 \longrightarrow 0
\end{equation}
and
\begin{equation}
0\longrightarrow {\bf C}\longrightarrow \Omega^0 \longrightarrow Z^1 \longrightarrow 0
\end{equation}
This implies that for $k>0$: $H^k(Z^2) = H^{k+1}(Z^1) = H^{k+2}({\bf C}) = \Lambda^{k+2}{\bf C}^d$.
Therefore, we obtain from (\ref{LongExactSequenceForMaxwell}):
\begin{align}
   H^0({\rm Maxwell}) = \;& \Lambda^2{\bf C}^{d}\;:\; f_{[mn]}dx^m\wedge dx^n
\\   
H^1({\rm Maxwell}) = \;& {\bf C}^{d} \oplus \Lambda^3{\bf C}^{d}\;:\;
c_kf_l dx^k\wedge dx^l\;\mbox{ \tt\small and }\; f_{[klm]}c^kdx^l\wedge dx^m
\label{H1Maxwell}\\    
H^2({\rm Maxwell}) = \;& {\bf C} \oplus \Lambda^4{\bf C}^{d}\;:\;
c_kc_ldx^k\wedge dx^l\;\mbox{ \tt\small and }\; f_{[ijkl]}c^ic^jdx^k\wedge dx^l
\label{H2Maxwell}\\  
H^{n>2}({\rm Maxwell}) = \;& \Lambda^{n+2}{\bf C}^d\;:\; f_{[j_1\ldots j_{n+2}]} c^{j_1}c^{j_2}\cdots c^{j_n}
dx^{j_{n+1}}\wedge dx^{j_{n+2}} 
\end{align}
Notice that all these cohomology classes are represented by the {\em constant}
field strength. In other words, the {\em dilatation symmetry} $x^m{\partial \over\partial x^m}$ acts as
zero in cohomology.

\subsection{Cohomology of ${\bf R}^{10}$ with values in solutions of Dirac equations}
Let $\cal D$ denote the space of solutions of the Dirac equations, and $\cal S$ the space
of chiral-spinor-valued functions, and ${\cal S}^*$ the antichiral-spinor-valued
functions. There is a short exact sequence:
\begin{equation}
   0\rightarrow {\cal D} \xrightarrow{\subset} {\cal S} 
\xrightarrow{\Gamma^m{\partial\over\partial x^m}} {\cal S}^* \rightarrow 0
\end{equation}
This leads to the long exact sequence of the cohomologies:
\begin{align}
   0\rightarrow {\bf C}^{16} \rightarrow {\bf C}^{16} \xrightarrow{0} {\bf C}^{16}
\rightarrow H^1({\cal D}) \rightarrow 0 \rightarrow 0 
\rightarrow H^2({\cal D}) \rightarrow 0 \rightarrow \cdots
\end{align}
Therefore:
\begin{align}
   H^0({\rm Dirac}) = \;& {\bf C}^{16} \;:\;\mbox{ \small\tt constant spinors}
\\  
H^1({\rm Dirac}) = \;& {\bf C}^{16}\;:\; \hat{c}\Psi \mbox{ \tt\small where }\Psi
\mbox{ \tt\small is constant}
\\   
H^{n>1}({\rm Dirac}) = \;& 0
\end{align}
 
\section{Zeroth cohomology of the tensor product of two classical electrodynamics}\label{sec:ZerothCohomology}
This is the direct sum:
\begin{align}
&   H^0({\rm Maxw}_L\otimes {\rm Maxw}_R) \oplus 
   H^0({\rm Dirac}_L\otimes {\rm Dirac}_R) \;\oplus
\nonumber \\  
\oplus \;&   H^0({\rm Maxw}_L\otimes {\rm Dirac}_R) \oplus
   H^0({\rm Dirac}_L\otimes {\rm Maxw}_R)
\end{align}
The space $H^0({\rm SMaxw}_L\otimes {\rm SMaxw}_R)$ can be thought of as the space of 
functions:
\begin{align}
   F_{[mn]\;;\;[pq]}(x )
\end{align}
satisfying:
\begin{align}
   \partial_{[k}F_{mn]\;;\;[pq]} \;&= 0
   \label{LeftJacobiOnFx}
   \\  
   F_{[mn]\;;\;[pq}\stackrel{\leftarrow}{\partial}_{r]} \;&= 0
   \label{RightJacobiOnFx}
   \\ 
   \partial^mF_{[mn]\;;\;[pq]} \;& = 0
   \label{LeftEqMOnFx}
   \\   
   F_{[mn]\;;\;[pq]}\stackrel{\leftarrow}{\partial}^q \;& = 0
   \label{RightEqMOnFx}
\end{align}
Eqs. (\ref{LeftJacobiOnFx})  and (\ref{RightJacobiOnFx}) together imply that:
\begin{align}
   F_{[mn\;;\;pq]} =&\; \mbox{const}
   \label{ConstantFAntisymm}
   \\   
   g^{mp}g^{nq}F_{[mn]\;;\;[pq]} =&\; \mbox{const}
   \label{ConstantDoubleTrace}
\end{align}
We can write:
\begin{equation}
   F_{[mn]\;;\;[pq]} \;= \partial_{[m}A^L_{n]\;;\;[pq]} =
   A^R_{[mn]\;;\;[p}\stackrel{\leftarrow}{\partial}_{q]}
\end{equation}
A consequence of Eqs. (\ref{LeftJacobiOnFx}), (\ref{RightJacobiOnFx}), (\ref{LeftEqMOnFx}), (\ref{RightEqMOnFx}) is the existence of
$\phi^R_q$ and $\phi^L_m$ such that:
\begin{align}
   \partial^mA^R_{[mn]\;;\;p} \;&=\partial_p\phi^R_n
\label{PhiRDef}
\\   
A^L_{m\;;\;[pq]}\stackrel{\leftarrow}{\partial}^q \;&=\partial_m\phi^L_p
\label{PhiLDef}
\end{align}
This implies:
\begin{align}
   g^{np}F_{[mn]\;;\;[pq]}(x) = 
   {1\over 2} \partial_q\left(
      g^{np}A^R_{[mn]\;;\;p} + \phi^R_m
   \right) = 
   {1\over 2}\partial_m\left(
      g^{np}A^L_{n\;;\;[pq]} + \phi^L_q
   \right)
\end{align}
Let us denote:
\begin{align}
   B^R_m \;&= g^{np}A^R_{[mn]\;;\;p} + \phi^R_m\;,\;\;
\\     
   B^L_q \;&= g^{np}A^L_{n\;;\;[pq]} + \phi^L_q
\end{align}
In particular:
\begin{align}
   \partial_m B^L_q = \partial_q B^R_m
\end{align}
Although $A^R_{[mn];p}$ and $\phi_n^R$ are only defined by (\ref{PhiRDef}) up to:
\begin{align}
 A^R_{[mn];p} &\mapsto A^R_{[mn];p} + \partial_p \chi^R_{mn}
 \\   
 \phi_n^R &\mapsto \phi_n^R + \partial^m \chi_{mn}\;,
\end{align}
this ambiguity does not affect the definition of $B_m^R$ (and similarly $B_q^L$).
Notice that:
\begin{align}
   \partial_{[m}B^R_{n]} =\;- \partial_{[p}B^L_{q]} =\;& \mbox{const}   
   \\ 
   \partial^p B^L_p = \;\partial^p B^R_p =\;& \mbox{const}
\end{align}
Let us denote:
\begin{align}
   B^L_m \pm B^R_m = A^{\pm}_m
\end{align}
Then:
\begin{align}
   \partial_{[q}A^{+}_{m]} \;& = 0 
\\    
\partial_{(q}A^{-}_{m)} \;& = 0
\label{EqAMinus}
\end{align}
The physical meaning of $A_m^{\pm}$ will be explained in Section \ref{sec:calE02}.

\section{First cohomology of the tensor product of two classical electrodynamics}\label{sec:CohomologyOfTensorProduct} 
Having computed the cohomology of $Q_{\rm Lie}$ with values in Maxwell and Dirac 
solutions, we will now use it to compute the cohomology with values in the
tensor product ${\rm SMaxw}_L\otimes {\rm SMaxw}_R$. Again, we will use some spectral sequence.
In order to distinguish it from the spectral sequence of Section \ref{sec:TypeIIBvsMaxwell}, we will 
use the notation\footnote{Unfortunately, because of certain limitations of LaTeX, we
can not afford similar notations for the differentials $d_r$} $E_r^{p,q}$ (that other one was denoted ${\cal E}_r^{p,q}$).
\subsection{Dirac-Dirac sector}\label{sec:DiracDiracSector}
\subsubsection{Spectral sequence $E_r^{p,q}$}
The following group is part of the ghost number 3 cohomology:
\begin{align}
H^1\left(Q_{\rm Lie}\;,\; {\rm Dirac}\otimes{\rm Dirac}\right)
\end{align}
In this section we will calculate this cohomology group.

The differential $Q_{\rm Lie}$ is realized on the space of bispinors $P^{\alpha\dot{\beta}}(x_L,x_R,c)$ 
satisfying:
\begin{align}
{\partial\over\partial x_L^m}\Gamma^m_{\alpha\alpha'}
P^{\alpha'\dot{\beta}}(x_L,x_R,c) = \;& 0
\\    
{\partial\over\partial x_R^m}P^{\alpha\dot{\beta}'}(x_L,x_R,c)
\Gamma^m_{\dot{\beta}'\dot{\beta}} =\;& 0
\end{align}
The differential $Q_{\rm Lie}$ acts as follows:
\begin{equation}
Q_{\rm Lie} P^{\alpha\dot{\beta}} = 
c^m\left({\partial\over\partial x_L^m} - {\partial\over\partial x_R^m}\right)
P^{\alpha\dot{\beta}} 
\end{equation}
Let us introduce the filtration by the degree $N$:
\begin{equation}
   N = {1\over 2}\left(
c {\partial\over\partial c} + x_L {\partial\over\partial x_L} 
- x_R{\partial\over\partial x_R}
\right)
\end{equation}
Then $c^m{\partial\over\partial x_L^m}$ is the leading (of degree zero) term in $Q_{\rm Lie}$ and $-c^m{\partial\over\partial x_R^m}$ is 
subleading (of degree one). Let us calculate the cohomology of $Q_{\rm Lie}$ using the
spectral sequence of this filtration. The first page $E_1^{p,q}$ is: 
\begin{align}
 E_1^{p,q} =\;&   H^{p+q}\left(c^m{\partial\over\partial x_L^m}\;,\; 
      {F^p({\rm Dirac}\otimes {\rm Dirac})
        \over 
        F^{p+1}({\rm Dirac}\otimes {\rm Dirac})}\right)
\\     
d_1 = \;& -c^m{\partial\over\partial x_R^m} \;:\;\; 
E_1^{p\;,\;q} \longrightarrow E_1^{p+1\;,\;q}
\end{align}
where $F^p$ consists of polynomials with $N\geq p$. Schematically, $E_1^{p,q}$ consists of
expressions of the form
\begin{equation}
  P^{\alpha\dot{\beta}} = [c^{p+q} x_L^{n+p} x_R^{n+q}]
\end{equation}
satisfying both left and right Dirac equations, representing the cohomology of 
$c^m{\partial\over\partial x_L^m}$. Just to remember:
\begin{equation}
   E_1^{{1\over 2}(\#c + \#x_L - \#x_R)\,,\,{1\over 2}(\#c + \#x_R - \#x_L)}
\end{equation}
where $\#x$ means ``degree in $x$``.

Because of Section \ref{sec:CohomologyInMaxwell}, the cohomology of $c^m{\partial\over\partial x_L^m}$ is localized on  
$n+p=0$, and either $p+q=0$ or $p+q=1$. This means that the only 
nontrivial components of $E_1^{p,q}$ are the ones represented by the following 
expressions:
\begin{align}
   E_1^{-m,m} \;:\;\;& P\langle x_R^{\otimes 2m}\rangle
\label{P}
\\    
E_1^{-m+1,m} \;:\;\;& \hat{c}R\langle x_R^{\otimes (2m-1)} \rangle
\label{cR}
\end{align}
Here, as usual, we denote $\hat{c} = c^m\Gamma_m$.

The only nontrivial differential is $d_1\;:\; E_1^{-m,m} \to E_1^{-m+1,m}$. The cohomology
of this differential is $E_2^{p,q}$. Notice that $d_2 = 0$. Indeed, the construction of 
$d_2:\; E_2^{p,q} \to E_2^{p+2,q-1}$ involves the inversion of $c^m{\partial\over\partial x_L^m}$ and therefore any 
expression in the image of $d_2$ is necessarily in the image of $x^m_L{\partial\over\partial x_L^m}$. But
$x^m_L{\partial\over\partial x_L^m}$ acts as zero on $E_1$ and therefore also on $E_2$.

Therefore our spectral sequence converges at the second page: $E_2 = E_{\infty}$.

\subsubsection{The image of $d_1(E_1^{-m,m})$}\label{sec:DefineR}
The condition that the cohomology class of an expression of the form (\ref{cR})
is cancelled by the $d_1$ of an expression of the form (\ref{P}) is:
\begin{align}\label{TrivialityOfR}
    R = \;& - {1\over 10}\Gamma^m{\partial\over\partial x_R^m}P
\\   
\mbox{\tt\small with } & {\partial\over\partial x_R^m}P\Gamma^m = 0
\label{DiracEquationOnP}
\end{align}
Indeed, for any $P(x_R)$ solving the right Dirac equation (\ref{DiracEquationOnP}) we can 
tautologically write:
\begin{equation}\label{PresentationOfR}
\hat{c}R = - c^m{\partial\over\partial x^m_R}P 
+ c^n{\partial\over\partial x_L^n} \left(
   \widehat{x}_L R + x_L^m{\partial\over\partial x_R^m} P
\right)
\end{equation}
Then (\ref{TrivialityOfR}) is the necessary and sufficient condition that $\Psi :=
   \widehat{x}_L R + x_L^m{\partial\over\partial x_R^m} P$
satisfies both ${\partial\over\partial x_R^m}\Psi\Gamma^m = 0$ {\em and} $\Gamma^m{\partial\over\partial x_L^m}\Psi = 0$. (And, moreover, any
presentation of $\hat{c}R$ as $-c^m{\partial\over\partial x_R^m}P$ plus $c^m{\partial\over\partial x_L^m}\mbox{(smth)}$ will necessarily be of the
form (\ref{PresentationOfR}).)

\paragraph     {Comment} Those $P$ which satisfy $\Gamma^m {\partial\over\partial x_R^m} P = 0$ are in the kernel
of $d_1$, and therefore they form $E_2^{-m,m}$. They are in the ghost number two 
cohomology (the Ramond-Ramond fields). We have previously explained that $d_2$ is 
zero; if it were not zero, it would have killed the ghost number two cohomology.

\vspace{20pt}

\noindent
Notice that any $P$ satisfying (\ref{TrivialityOfR}) and (\ref{DiracEquationOnP}) is automatically harmonic:
$\Delta P = 0$, therefore (\ref{TrivialityOfR}) and (\ref{DiracEquationOnP}) imply that $R$ satisfies the left Dirac 
equation:
\begin{equation}\label{RSatisfiesLeftDirac}
   \Gamma^m{\partial\over\partial x_R^m} R = 0
\end{equation}
This means that:
\begin{equation}\label{DiracOperatorIsObstacle}
 \Gamma^m{\partial\over\partial x_R^m}R \;\;
\mbox{ \tt\small is an obstacle for the triviality of } R
\end{equation}
In the rest of this section we will prove that this is the only obstacle, 
{\it i.e.} any $R$ satisfying (\ref{RSatisfiesLeftDirac}) can be represented as (\ref{TrivialityOfR}), (\ref{DiracEquationOnP}). 

\subsubsection{Proof that (\ref{DiracOperatorIsObstacle}) is the only obstacle to the triviality of $R$}
In this section we will prove that if $R$ is a polynomial of nonzero degree
({\it i.e.} not a constant), than (\ref{DiracOperatorIsObstacle}) is the only obstacle to the triviality 
of $R$.

Notice that it is always possible to solve for $P$ to satisfy (\ref{TrivialityOfR}), but $P$ will
not necessarily satisfy (\ref{DiracEquationOnP}). But if the Dirac equation (\ref{RSatisfiesLeftDirac}) is satisfied, 
then we have:
\begin{align}
\Gamma^m\;\partial_m\partial_nP\;\Gamma_n =\; & 0
\label{DLDRPIsZero}\\     
\Delta P = \;& 0
\label{PIsHarmonic}
\end{align}
We will now prove that (\ref{DLDRPIsZero}) and (\ref{PIsHarmonic}) imply that exist $P_L$ and $P_R$ such that:
\begin{align}
   P \;& = P_L + P_R
\\    
\mbox{\tt\small where }\;& \Gamma^m {\partial\over\partial x_R^m} P_L = 0
\;\mbox{ \tt\small and }\; {\partial\over\partial x_R^m} P_R \Gamma^m = 0
\end{align}
This implies that $P$ can be chosen to satisfy the right Dirac equation, and
therefore $R$ is in the image of $d_1$.

\paragraph     {Proof.} 
Let us switch from the bispinor notations to the forms notations. The left
Dirac operator corresponds to ${\cal D}_L = d + \delta$ while the right Dirac operator is 
${\cal D}_R = (-1)^{F+1}(d - \delta)$. Eq. (\ref{PIsHarmonic}) implies that $(\delta d + d\delta)P =0$ while Eq. (\ref{DLDRPIsZero}) 
implies that $(\delta d - d\delta)P =0$. Therefore we have:
\begin{equation}\label{dDeltaPIsZero}
   d\delta P = \delta d P  = 0
\end{equation}
We will now prove that under the condition (\ref{dDeltaPIsZero}) exist $P_L$ and $P_R$ such that:
\begin{align}
   P \;& = P_L + P_R
\nonumber \\    
{\cal D}_L P_L \;& = {\cal D}_R P_R = 0
\label{PLAndPR}
\end{align}
It is useful to keep in mind the cohomology of the de Rham $d$ on harmonic forms
is:
\begin{equation}
   H^0(d,\mbox{ker} \Delta) = H^1(d,\mbox{ker} \Delta) = {\bf C} \quad,\quad
   H^{>1}(d,\mbox{ker} \Delta) = 0
\end{equation}
(the $H^1(d,\mbox{ker} \Delta)$ is generated by $x^mdx^m$).

\paragraph     {Case when $P$ is a 5-form}
In this case we will write $P^{(5)}$ instead of $P$.
Since $d\delta P^{(5)} = 0$, exists a harmonic 3-form $P^{(3)}$ such that:
\begin{equation}
   \delta P^{(5)} = d P^{(3)}
\end{equation}
Similarly, as $\delta d P^{(5)} = 0$, exists a harmonic 7-form $P^{(7)}$ such that:
\begin{equation}
   dP^{(5)} = \delta P^{(7)}
\end{equation}
Furthermore, there exist harmonic $P^{(1)}$ and $P^{(9)}$ such that:
\begin{equation}
   \delta P^{(3)} = d P^{(1)} \;\mbox{ \tt\small and }\; dP^{(7)} = \delta P^{(9)}
\end{equation}
This implies that $\delta P^{(1)} = 0$ and therefore exists a harmonic form $S^{(2)}$ such that
$P^{(1)} = \delta S^{(2)}$.  Similarly, $P^{(9)} = d S^{(8)}$. Therefore the following $P_L$ and $P_R$
satisfy (\ref{PLAndPR}):
\begin{align}
   P_L \;& = {1\over 2}\left(
P^{(5)} - (P^{(3)} + d S^{(2)}) - (P^{(7)} + \delta S^{(8)})
\right)
\\    
P_R \;& = {1\over 2}\left(
P^{(5)} + (P^{(3)} + d S^{(2)}) + (P^{(7)} + \delta S^{(8)})
\right)
\end{align}

\paragraph     {Case when $P$ is a 3-form plus 7-form}
The 7-form part of $P$ is related to the 3-form part by the condition that $P$ 
is self-dual. In this case we will write $P^{(3)} + P^{(7)}$ instead of $P$. (This $P^{(3)}$ 
has nothing to do with the $P^{(3)}$ of the previous paragraph.) Since $d\delta P^{(3)} = 0$
and $\delta\delta P^{(3)} = 0$, exists harmonic $P^{(1)}$ such that:
\begin{equation}
 \delta P^{(3)} = d P^{(1)}  
\end{equation}
This implies that $\delta P^{(1)} = 0$. Similarly, exists a harmonic $P^{(5)}$ such that:
\begin{equation}
   dP^{(3)} = \delta P^{(5)}
\end{equation}
This automatically implies:
\begin{equation}
   dP^{(5)} = \delta P^{(7)}
\end{equation}
Also exists a harmonic $P^{(9)}$ such that:
\begin{equation}
   \delta P^{(9)} = d P^{(7)} \mbox{ \tt\small and } dP^{(9)} = 0
\end{equation}
We take:
\begin{align}
   P_L \;& = {1\over 2}\left(
- P^{(1)} + P^{(3)} - P^{(5)} + P^{(7)} - P^{(9)}
\right)
\\     
P_R \;& = {1\over 2}\left(
 P^{(1)} + P^{(3)} + P^{(5)} + P^{(7)} + P^{(9)}
\right)
\end{align}

\paragraph     {Case when $P$ is a 1-form plus a 9-form}
Now suppose that $P = P^{(1)} + P^{(9)}$.  Let us first assume that the degree of $P$
is more than 1. We have:
\begin{equation}
   d\delta P^{(1)} =0 \;\Rightarrow\; \delta P^{(1)} = 0 
\;\Rightarrow\; P^{(1)} = \delta S^{(2)}
\end{equation}
Similarly $P^{(9)} = dS^{(8)}$. Now we have:
\begin{align}
   P_L \;&= {1\over 2} (\delta + d) S^{(2)} + {1\over 2}(d + \delta) S^{(8)}
\\   
P_R \;&= {1\over 2} (\delta - d) S^{(2)} + {1\over 2}(d - \delta) S^{(8)}
\end{align}
Now consider the case when the degree of $P$ is one, {\it i.e.} $P$ is linear 
in $x$. In this case we can have $\delta P^{(1)} = \mbox{const}$. This corresponds to the $R$
of (\ref{TrivialityOfR}) a constant proportional to unit matrix. The corresonding element
of $H^1(Q_{\rm Lie}\;,\;{\rm Dirac}\otimes {\rm Dirac})$ is:
\begin{equation}
   (\theta_L\Gamma^m\lambda_L)\;
   (\theta_L\Gamma_m\hat{c}\Gamma_n\theta_R)\;
   (\lambda_R\Gamma^n\theta_R)
\end{equation}
It corresponds to the following ghost number three vertex:
\begin{equation}\label{VertexGhostNumberThreeDimFour}
      (\theta_L\Gamma^m\lambda_L)(\theta_L\Gamma^p\lambda_L)\;
   (\theta_L\Gamma_m\Gamma_p\Gamma_n\theta_R)\;
   (\lambda_R\Gamma^n\theta_R)
\end{equation}

\paragraph     {Conclusion}
We conclude that the main obstacle for (\ref{cR}) to be trivial is $\Gamma^m\partial_mR\neq 0$. 
(And besides that, there is also a case when $R$ is a constant times a unit 
matrix, which results in a nontrivial vertex (\ref{VertexGhostNumberThreeDimFour}).)
If $\Gamma^m\partial_mR\neq 0$, then there is a nontrivial cohomology class of the form:
\begin{align}
   c^n\Gamma_n R 
   + r_1[cx_L x_R^{(2m-2)}]
   + r_2[c x_L^2 x_R^{(2m-3)}]
   + \ldots + r_{2m-1}[c x_L^{(2m-1)}]
\label{SeriesForRepresentativeH1DiracDirac}
\end{align}
Indeed, acting on the leading term $c^n\Gamma_n R$ with $-c^m{\partial\over\partial x_R^m}$ we get an expression 
of the form $[c^2 x_R^{2m-2}]$, which does not depend on $x_L$ and therefore is 
annihilated by $c^m{\partial\over\partial x_L^m}$. But since $H^2\left(c{\partial\over\partial x}\;,\;{\rm Dirac}\right) = 0$, this expression is 
automatically of the form $c^m{\partial\over\partial x_L^m}[cx_Lx_R^{(2m-2)}]$. Continuing this process we
get (\ref{SeriesForRepresentativeH1DiracDirac}). 

\subsubsection{Proof that $V_3$ of Eq. (\ref{V3}) is BRST nontrivial}
Let us consider the ghost number three vertex $V_3$ given by  Eq. (\ref{V3}), and 
expand it in the Taylor series in $x$ and $\theta$. We assign to $x$  degree $1$ and to
$\lambda$ and $\theta$ degree $1/2$. The BRST operator preserves this degree. In particular, 
every term in the expansion is a BRST-closed {\em polynomial} of $x,\lambda,\theta$. It is 
enough to prove the nontriviality term by term. Let us consider the extended 
space $(x_L,x_R,\lambda_L,\lambda_R,\theta_L,\theta_R)$. In this extended space, we get:
\begin{equation}
   V_3 = (Q_L + Q_R)\left((a\cdot (x_L - x_R)) V_2\right)
\end{equation}
The corresponding element of $H^1\left(c\left({\partial\over\partial x_L} - {\partial\over\partial x_R}\right)\;,\; {\rm Dirac}\otimes {\rm Dirac}\right)$ is given by:
\begin{equation}\label{ADotCPExp}
(a\cdot c) P e^{k(x_L + x_R)}   
\end{equation}
Consider the expansion in powers of $x_L$. The leading term is $(a\cdot c) Pe^{kx_R}$. 
We observe:

\begin{align}
   \left(10(c\cdot a) - \hat{c}\hat{a}\right)  = 
c{\partial\over\partial x_L}
(4\hat{x}_L \hat{a} + 5\hat{a}\hat{x}_L)
\end{align}
and $(4\hat{x}_L \hat{a} + 5\hat{a}\hat{x}_L)Pe^{kx_R}$ satisfies the left Dirac equation. Therefore (\ref{ADotCPExp}) is
equivalent to ${1\over 10} \hat{c}\hat{a} Pe^{kx_R}$. Comparing this with (\ref{SeriesForRepresentativeH1DiracDirac}), we get:

\begin{align}
   R \;&= {1\over 10}\hat{a}Pe^{kx_R}
\\    
\Gamma^m{\partial\over\partial x_R^m} R \;&=  {1\over 5}(a\cdot k) Pe^{kx_R}
\neq 0
\end{align}
Then (\ref{DiracOperatorIsObstacle}) implies that $V_3$ represents a nontrivial cohomology class.

\paragraph     {Ghost number three vertex of \cite{Mikhailov:2012uh}} can be obtained as the first order
of expansion of (\ref{ADotCPExp}) in powers of $x$. Indeed, at the first order of the 
$x$-expansion $R = {1\over 10}\hat{a}(k\cdot x_R) P$. Notice that the expression:
\begin{equation}
\left(\hat{a}(k\cdot x_R) - {1\over 5} \hat{x}_R(k\cdot a)\right)P
\end{equation}
satisfies the left Dirac equation (we use $\hat{k}P = 0$). Therefore
$R = {1\over 10}\hat{a}(k\cdot x_R) P$ is equivalent to $R = {1\over 50} \hat{x}_R(k\cdot a)P$.  Therefore the leading
term of the $x$-linear part of (\ref{ADotCPExp}) is equivalent to ${1\over 50} \hat{c}\hat{x}_R(k\cdot a) P$ which is the
leading term of the vertex constructed in \cite{Mikhailov:2012uh}.

\subsection{Maxwell-Maxwell sector}
In this section we will compute the cohomology of $c^m \left({\partial\over\partial x_L^m} - {\partial\over\partial x_R^m}\right)$ on the 
solutions of bi-Maxwell equations. 

\subsubsection{Bi-Maxwell equations}
Solutions of bi-Maxwell equations  are defined as expressions of the form:
\begin{equation}\label{BiMaxwell}
dx_L^p\wedge dx_L^q\;
\left(
   {\partial\over \partial x_L^{[p}}{\cal A}(x_L,x_R)_{q]\;;\;[m}
   {\stackrel{\leftarrow}{\partial}\over\partial x_R^{n]}}
\right)\;
dx_R^m\wedge dx_R^n\;
\end{equation}
satisfying the left and right Maxwell equations:
\begin{align}
   {\partial\over\partial x_L^p}{\partial\over\partial x_L^{[p}}{\cal A}(x_L,x_R)_{q]\;;\;[m}{\stackrel{\leftarrow}{\partial}\over\partial x_R^{n]}} = 0
\\   
{\partial\over\partial x_L^{[p}}
{\cal A}(x_L,x_R)_{q]\;;\;[m}
{\stackrel{\leftarrow}{\partial}\over\partial x_R^{n]}}
{\stackrel{\leftarrow}{\partial}\over\partial x_R^{n}} = 0
\end{align}
Notice that we have left and right indices, separated with the semicolon. We
use the notations ${\stackrel{\leftarrow}{\partial}\over \partial x}$. The expression $\phi {\stackrel{\leftarrow}{\partial}\over \partial x}$ means the same as ${\partial\over\partial x}\phi$. The sole
purpose of such notations is to improve the readability of the formulas, as 
they allow us to naturally separate left and right indices.

\subsubsection{Spectral sequence $E_r^{p,q}$}
\paragraph     {Definition}
As in Section \ref{sec:DiracDiracSector}, we will use the filtration by the powers of $x_L$, {\it i.e.} 
treat $x_L$ as being small. The elements of $E_r^{p,q}$ are of the type: 
\begin{equation}\label{E1pqMaxwellMaxwell}
E_r^{p,q}\;:\;\;dx_L\wedge dx_L \;[c^{p+q}x_L^{n+p}x_R^{n+q}]\;dx_R\wedge dx_R + \ldots
\end{equation}
where $\ldots$ stands for terms of the type $dx_L\wedge dx_L \;[c^{p+q}x_L^{n+p+s}x_R^{n+q-s}]\;dx_R\wedge dx_R$
with $s>0$, which are factored out when we consider $F^p({\rm Maxwell}\otimes {\rm Maxwell})$ 
modulo $F^{p+1}({\rm Maxwell}\otimes {\rm Maxwell})$. For a polynomial element ${\cal A}_{q\;;\;m}$, of the 
total order $M$ in $x_L$ and $x_R$, there is an expansion in powers of $x_L$:
\begin{equation}\label{MaxwellMaxwellExpansion}
   {\cal A}(x_L,x_R) = {\cal A}^{(0)}_{q\;;\;m}(x_R) + 
   {\cal A}^{(1)}_{q\;;\;m}(x_L,x_R) + \ldots +
   {\cal A}^{(N)}_{q\;;\;m}(x_L)
\end{equation}
where ${\cal A}^{(0)}_{q\;;\;m}$ does not depend on $x_L$, ${\cal A}^{(1)}_{q\;;\;m}$ is linear in $x_L$, {\it etc.}.

\paragraph     {The structure of $E_2^{p,q}$}
The following is the most general (up to the $c{\partial\over\partial x_L}$-exact terms) ansatz for the 
leading term ${\cal A}^{(0)}_{q\;;\;m}$:
\begin{align}
{\cal A}^{(0)}_{q\;;\;m} = \;& \phantom{+\;}
c_{p} dx_L^p\wedge dx_L^q\; 
A(x_R)_{q\;;\;[m}\stackrel{\leftarrow}{\partial}_{n]}\;dx_R^m\wedge dx_R^n\; + 
\label{AnsatzMaxwellMaxwell} \\   
 \;&+ c^p dx_L^q \wedge dx_L^r\; B(x_R)_{pqr\;;\;[m}\stackrel{\leftarrow}{\partial}_{n]}\;dx_R^m\wedge dx_R^n\;
\nonumber \\     
\mbox{\small\tt with }\;& 
A(x_R)_{q\;;\;[m}\stackrel{\leftarrow}{\partial}_{n]}
\stackrel{\leftarrow}{\partial}_m = 0
\label{AnsatzMaxwellMaxwellEqm}\\   
& B(x_R)_{pqr\;;\;[m}\stackrel{\leftarrow}{\partial}_{n]}
\stackrel{\leftarrow}{\partial}_m = 0
\label{AnsatzMaxwellMaxwellEqmForB}\\   
& \partial^qA(x_R)_{q\;;\;[m}\stackrel{\leftarrow}{\partial}_{n]} = 0
\label{AnsatzMaxwellMaxwellUnobstructedA}\\
& \partial_{[p}B(x_R)_{qrs]\;;\;[m}\;\stackrel{\leftarrow}{\partial}_{n]} = 0
\label{AnsatzMaxwellMaxwellUnobstructedB}\\   
{\cal A}^{(0)}_{q\;;\;m}\;& \mbox{ \tt\small represents} 
\mbox{ \tt\small an element of }
E_2^{-{M-1\over 2},{M+1\over 2}} \mbox{ \tt\small , see (\ref{E1pqMaxwellMaxwell})}
\end{align}
Here $A(x_R)_{q\;;\;m}$ and $B(x_R)_{pqr\;;\;m} = B(x_R)_{[pqr]\;;\;m}$ are polynomials in $x_R$ of the
order $M$. They correspond to the two terms in (\ref{H1Maxwell}). Eqs. (\ref{AnsatzMaxwellMaxwellEqm}) and (\ref{AnsatzMaxwellMaxwellEqmForB}) 
enforce the right Maxwell equation. (The left Maxwell equation is automatically
satisfied because $A$ does not depend on $x_L$.) Eqs. (\ref{AnsatzMaxwellMaxwellUnobstructedA}) and (\ref{AnsatzMaxwellMaxwellUnobstructedB}) are the 
conditions for being in the kernel of $d_1$. In other words, those are the 
conditions for the existence of $A^{(1)}(c, x_L,x_R)_{p\;;\;m}$ linear in $x_L$ and $c$ such 
that:
\begin{align}
 c^r{\partial\over\partial x^r_R}
   \Big( &
      c_{p} dx_L^p\wedge dx_L^q\; A(x_R)_{q\;;\;[m}
      \stackrel{\leftarrow}{\partial}_{n]}\;dx_R^m\wedge dx_R^n\; +
\nonumber \\   
& + \; 
c^p dx_L^q \wedge dx_L^r\; B(x_R)_{pqr\;;\;[m}\stackrel{\leftarrow}{\partial}_{n]}\;dx_R^m\wedge dx_R^n\;
    \Big) \; =
   \nonumber \\    
=\;& c^r{\partial\over\partial x^r_L}
   \left(
      dx_L^p\wedge dx_L^q\;\partial_{[p} A^{(1)}(c,x_L,x_R)_{q]\;;\;[m}
      \stackrel{\leftarrow}{\partial}_{n]}\;dx_R^m\wedge dx_R^n\;
   \right)
\\   
\mbox{\tt\small and } &
A^{(1)}(x_L, x_R)_{q\;;\;[m}\stackrel{\leftarrow}{\partial\over \partial x_R^{n]}}
\stackrel{\leftarrow}{\partial\over\partial x_R^m} \; = \; 
{\partial\over\partial x_L^p}{\partial\over\partial x_L^{[p}}
A^{(1)}(x_L, x_R)_{q]\;;\;m}
\;= \;0
\end{align}
Eq. (\ref{AnsatzMaxwellMaxwellUnobstructedA}) is the vanishing of the obstacle proportional to the first term in 
(\ref{H2Maxwell}), and Eq. (\ref{AnsatzMaxwellMaxwellUnobstructedB}) is to avoid hitting the second term in (\ref{H2Maxwell}).

Remember that we are working in the polynomial sector, {\it i.e.} 
$B_{pqr\;;\;[m}(x_R)\stackrel{\leftarrow}{\partial}_{n]}$ is a homogeneous polynomial in $x_R$. Let us first assume that 
the degree of the polynomial is nonzero:
\begin{equation}
B_{pqr\;;\;[m}(x_R)\stackrel{\leftarrow}{\partial}_{n]} \;\neq\; \mbox{const}
\end{equation}
Then (\ref{AnsatzMaxwellMaxwellUnobstructedB}) implies that we can remove the term 
$c^p dx_L^q \wedge dx_L^r\; B(x_R)_{pqr\;;\;[m}\stackrel{\leftarrow}{\partial}_{n]}\;dx_R^m\wedge dx_R^n$, by adding to ${\cal A}^{(0)}$ an
element in $d_1(E_1^{-{M+1\over 2}, {M+1\over 2}})$. Indeed, this is equivalent to the existence of the
following two objects:
\begin{itemize}
\item $C(x_R)_{pq\;;\;m}$ satisfying $C(x_R)_{pq\;;\;[m}\stackrel{\leftarrow}{\partial}_{n]}\stackrel{\leftarrow}{\partial}_{m} = 0$ and
\item $G(x_L,x_R)_{pq\;;\;m}$ linear in $x_L$ satisfying left and right Maxwell equations:
{\small
\begin{align}
{\partial\over\partial x_L^{[p}}G(x_L,x_R)_{qr]\;;\;[m}\stackrel{\leftarrow}{\partial\over\partial x_R^{n]}} \;&= 0  
\nonumber \\   
{\partial\over\partial x_L^p} G(x_L,x_R)_{pq\;;\;[m}\stackrel{\leftarrow}{\partial\over\partial x_R^{n]}} \;&= 0 
\nonumber \\
G(x_L,x_R)_{pq\;;\;[m}\stackrel{\leftarrow}{\partial\over\partial x_R^{n]}}{\stackrel{\leftarrow}{\partial}\over\partial x_R^{m}} \;&= 0
\nonumber 
\end{align}
}
\end{itemize}
such that:
\begin{align}
   & c^p dx_L^q \wedge dx_L^r\; B(x_R)_{pqr\;;\;[m}\stackrel{\leftarrow}{\partial}_{n]}\;dx_R^m\wedge dx_R^n \; + O(x_L) =
\nonumber \\    
=\;\;\;\;& 
c^k\left({\partial\over\partial x^k_L} - {\partial\over\partial x^k_R}\right)\;
\Big(\;\;\; dx_L^p\wedge dx_L^q\;
C(x_R)_{pq\;;\;[m}\stackrel{\leftarrow}{\partial}_{n]}\;dx_R^m\wedge dx_R^n \; +
\nonumber \\  
&  
\phantom{c^k\left({\partial\over\partial x^k_R} - {\partial\over\partial x^k_R}\right)}
\;\;+\; dx_L^p\wedge dx_L^q\;
G(x_L,x_R)_{pq\;;\;[m}
{\stackrel{\leftarrow}{\partial}\over\partial x_R^{n]}}
\;dx_R^m\wedge dx_R^n\;\;
\Big)
\;+ 
\nonumber \\
\;& +
c_{p} dx_L^p\wedge dx_L^q\; 
\widetilde{A}(x_R)_{q\;;\;[m}\stackrel{\leftarrow}{\partial}_{n]}\;
dx_R^m\wedge dx_R^n\; + \;
 O(x_L) \;
\label{RemovingBTerm} 
\end{align}
Here $\widetilde{A}(x_R)_{q\;;\;[m}\stackrel{\leftarrow}{\partial}_{n]}$ is some correction to the $A(x_R)_{q\;;\;[m}\stackrel{\leftarrow}{\partial}_{n]}$ of (\ref{AnsatzMaxwellMaxwell}). (In
other words, when we gauge away the $B$-term, this leads to some change in the
$A$ term: $A\to A + \widetilde{A}$.) The existence of such  $C(x_R)_{pq\;;\;m}$ 
and  $G(x_L,x_R)_{pq\;;\;m}$ follows from (\ref{AnsatzMaxwellMaxwellUnobstructedB}) and the fact that $H^3(Q_{\rm Lie},{\rm Maxw})$ 
is zero in polinomials of the degree $>0$, 
in the following way\footnote{Notice that we are using the results about $H(Q_{\rm Lie}, {\rm Maxw})$ in two different ways. First, we use $H^1(Q_{\rm Lie},{\rm Maxw}_L)$ to argue that the leading term can be reduced to the form (\ref{AnsatzMaxwellMaxwell}). Then we use the {\em vanishing} of $H^3(Q_{\rm Lie}, {\rm Maxw}_R)$ in polynomials of high enough degree to remove the B-term by adding 
$d_1(\mbox{smth})$. }. 
Eq. (\ref{AnsatzMaxwellMaxwellUnobstructedB}) implies that exists $C(x_R)_{pq\;;\;m}$ satisfying the right Maxwell 
equation, such that:
\begin{equation}
   B_{pqr\;;\;[m}\stackrel{\leftarrow}{\partial}_{n]} 
= - \partial_{[p} C_{qr]\;;\;[m} \stackrel{\leftarrow}{\partial}_{n]} 
\end{equation}
Therefore, in computing the first line on the RHS of (\ref{RemovingBTerm}), the $c^k{\partial\over\partial x_L^k}$ gives 
zero as $C_{pq\;;\;m}$ does not depend on $x_L$, and when acting with $-c^k{\partial\over\partial x_R^k}$, we get:
\begin{equation}
   - c^k {\partial\over\partial x_R^k}\; dx_L^p\wedge dx_L^q\;
C(x_R)_{pq\;;\;[m}\stackrel{\leftarrow}{\partial}_{n]}\;dx_R^m\wedge dx_R^n \;
\end{equation}
This has to be understood as an element of $E_1^{-{M-1\over 2},{M+1\over 2}}$, {\it i.e.} modulo the
image of $c^k{\partial\over\partial x_L^k}$. This ambiguity is described by the second line on the RHS 
of (\ref{RemovingBTerm}), the term containing $G(x_L,x_R)_{pq\;;\;m}$. This term can be used to remove
the components other than those listed in Eq. (\ref{H1Maxwell}); the component ${\bf C}^d$
corresponds to $\widetilde{A}_{q\;;\;m}$, and the component $\Lambda^3 {\bf C}^d$ kills the B-term.

We conclude that we can get rid of the $B$-term in (\ref{AnsatzMaxwellMaxwell}) by adding to ${\cal A}$ an 
element in $d_1(E_1^{-{M+1\over 2}, {M+1\over 2}})$.

Now let us consider the case when $B_{pqr\;;\;[m}(x_R)\stackrel{\leftarrow}{\partial}_{n]}$ is constant:
\begin{equation}
B_{pqr\;;\;[m}(x_R)\stackrel{\leftarrow}{\partial}_{n]} = \mbox{const}
\end{equation}
Consider the total antisymmetrization:
\begin{equation}\label{DefBpqrmn}
   {\cal B}_{[pqrmn]} = B_{[pqr\;;\;m}(x_R)\stackrel{\leftarrow}{\partial}_{n]} 
\end{equation}
In this case the $B$-term in (\ref{AnsatzMaxwellMaxwell}) cannot be gauged away, as ${\cal B}_{[pqrmn]}$ represents
a nontrivial cohmology class of $H^3({\rm Maxw}) = \Lambda^5{\bf C}^{10}$. However, we will show
in Section \ref{sec:GhostNumberFour} that this is cancelled by the $d_2\;:\;{\cal E}^{1,2}_2 \to {\cal E}^{3,1}_2$. In other words,
for our ansatz to survive on ${\cal E}_3^{1,2}$ we need to put ${\cal B}_{[pqrmn]}$ to zero:
\begin{equation}
   {\cal B}_{[pqrmn]}  = 0
\end{equation}

\subsubsection{Double field strength}\label{sec:DoubleFieldStrength}
Let us therefore assume that $B(x_R)_{pqr\;;\;m}=0$.  Can the remaining $A$-term also
be in the image of $d_1$? Let us define the double field strength $F_{[pq];[mn]}$ as 
follows:
\begin{equation}\label{DefineDoubleFieldStrength}
   F_{[pq];[mn]} = \partial_{[p}A_{q]\;;\;[m}\stackrel{\leftarrow}{\partial}_{n]}
\end{equation}
This double field strength has the following properties:
\begin{align}
F_{[pq\;;\;mn]} \;& = 0 \mbox{ \tt\small (total antisymmetrization)}
\\   
\partial_{[p}F_{qr]\;;\;mn} \;&= 0
\\    
F_{qr\;;\;[mn}\stackrel{\leftarrow}{\partial}_{k]} \;& = 0
\\   
\partial^p F_{[pq];[mn]} \;&= 0
\\    
F_{[pq];[mn]} \stackrel{\leftarrow}{\partial}^m \;&= 0
\\   
\Delta F_{[pq];[mn]} = 0
\end{align}

\subsubsection{Double field strength is the obstacle to triviality}
We will now show that $\cal A$ is trivial iff $F_{pq\;;\;mn} = 0$. 

We have to understand under which conditions the class with the leading term 
(\ref{AnsatzMaxwellMaxwell}) is trivial, {\it i.e.} can be obtained by acting with  $c^m \left({\partial\over\partial x_L^m} - {\partial\over\partial x_R^m}\right)$ 
on something:
\begin{align}
   \;& c_{[p} dx_L^p\wedge dx_L^q\; A(x_R)_{q]\;;\;[m}\stackrel{\leftarrow}{\partial}_{n]}\;dx_R^m\wedge dx_R^n\; + \ldots =
\\    
=\;&
c^j\left(
{\partial\over\partial x_L^j} - {\partial\over\partial x_R^j}\right)\;
\left(
dx_L^p\wedge dx_L^q\;
W(x_R)_{pq\;;\;[m}{\stackrel{\leftarrow}{\partial}\over \partial x_R^{n]}}\;
dx_R^m\wedge dx_R^n\; + \ldots
\right)
\nonumber
\end{align}
This property is equivalent to the existence of $W(x_R)_{pq\;;\;m}$ satisfying:
\begin{align}
   A_{q \;;\; [m} \stackrel{\leftarrow}{\partial}_{n]}
\; = \;& 
\partial^{p} W_{pq \;;\; [m} 
\stackrel{\leftarrow}{\partial}_{n]}
\label{AFromW}
\\    
\mbox{\tt\small with } \;&
\partial_{[p}W_{qr]\;;\;[m}\;\stackrel{\leftarrow}{\partial}_{n]} = 0 
\label{WIsLeftClosed}\\  
\mbox{\tt\small and }\;& 
W_{pq\;;\;[m} \stackrel{\leftarrow}{\partial}_{n]} \stackrel{\leftarrow}{\partial}_{n} = 0  
\label{RightMaxwellOnW}
\end{align}
Notice that the ghost number two vertices correspond to $W_{pq\;;\;m}$ satisfying
(\ref{WIsLeftClosed}), (\ref{RightMaxwellOnW}) and $\partial^{p} W_{pq \;;\; [m} 
\stackrel{\leftarrow}{\partial}_{n]} = 0$ instead of (\ref{AFromW}).

\vspace{20pt}

\noindent If $A$ can be expressed through $W$ as in (\ref{AFromW}), then we have:
\begin{align}
F_{pq \;;\;  mn} = \;&
\partial_{[p}A_{q]\;;\;[m}\stackrel{\leftarrow}{\partial}_{n]} =
-\;\partial_{[p}\partial^rW_{q]r\;;\;[m}\;\stackrel{\leftarrow}{\partial}_{n]} = 
{1\over 2}\partial_r\partial^rW_{pq\;;\;[m}\;\stackrel{\leftarrow}{\partial}_{n]} 
\;= 0
\end{align}
This means that: 
\begin{equation}
F_{pq\;;\;mn}\neq 0 
\mbox{ \small\tt is an obstacle to the triviality of ${\cal A}$ }
\end{equation} 
We will now prove that this is the only obstacle. In other words, 
if $F_{pq\;;\;mn} = 0$, then (\ref{MaxwellMaxwellExpansion}) is cohomologically trivial. 

Let ${\cal M}$ be the space of polynomial expressions of the form:
\begin{equation}\label{SpaceWhereDLPlusDeltaLActs}
\Phi(dx_L,x)_{[mn]} 
\;\; \mbox{\tt\small satisfying } \;
\Phi_{[mn}
\stackrel{\leftarrow}{\partial}_{k]}  = 0 
\;\; \mbox{\tt\small and } \;
\Phi_{mn}\stackrel{\leftarrow}{\partial}_n = 0
\end{equation}
Let ${\cal M}^N$ be the subspace of ${\cal M}$ consisting of polynomials of the order $N$ in 
$x$, {\it i.e.}  $x^p{\partial\over\partial x^p}\Phi = N\Phi$. Notice that such $\Phi_{mn}$ are automatically 
harmonic. The operator $d_L + \delta_L$ acts on such expressions, and is nilpotent: 
\begin{equation}
  \ldots \longrightarrow 
{\cal M}^N \xrightarrow{d_L + \delta_L} {\cal M}^{N-1} \longrightarrow \ldots
\end{equation}
\paragraph     {Lemma} 
\begin{equation}\label{VanishingOfCohomologyDPlusDelta}
   H^N(d_L + \delta_L\;,\;{\cal M}) = H^N(d_L\;,\;{\cal M}) 
= H^N(\delta_L\;,\;{\cal M}) = 0 \;\;\mbox{\tt\small for }\; N > 0
\end{equation}
Indeed, $d_L$ is acyclic, as $H^N(d_L)$ is the same as already computed in Section 
\ref{sec:CohomologyInMaxwell} cohomology of the translations algebra on the solutions of the Maxwell 
equations, and it is zero for $N>0$. This means that it is always possible to
gauge away the term with the highest number of $dx_L$, and therefore the 
cohomology of $d_L + \delta_L$ is zero. The proof of $H^N(\delta_L) = 0$ is identical to the 
proof of $H^N(d_L)=0$ after applying the Hodge dual operation on the $c$ ghosts.

Eq. (\ref{AnsatzMaxwellMaxwellEqm}) implies that the expression $dx_L^p\;A_{p\;;\;[m} \stackrel{\leftarrow}{\partial}_{n]}$ belongs to $\cal M$. Eq. (\ref{AnsatzMaxwellMaxwellUnobstructedA})
implies that it is annihilated by $\delta_L$. Since $H^N(\delta_L) = 0$, exists $\Phi^{(2)}\in {\cal M}$ 
such that:
\begin{equation}\label{FromAToPhi2}
   dx_L^p \;  A_{p\;;\;[m}\stackrel{\leftarrow}{\partial}_{n]} = 
\delta_L \left( dx_L^p\wedge dx_L^q\;\Phi^{(2)}_{pq\;;\;mn} \right)
\end{equation}
Now suppose that $F_{pq\;;\;mn} = 0$. This implies that we can find $\Phi^{(4)}$, $\Phi^{(6)}$, $\Phi^{(8)}$ 
and $\Phi^{(10)}$ (all elements of ${\cal M}$) satisfying:
\begin{equation}\label{ChainOfPhis}
   dx_L^p \;  A_{p\;;\;[m}\stackrel{\leftarrow}{\partial}_{n]} = 
(\delta_L + d_L) \left( \Phi^{(2)}_{mn} + \Phi^{(4)}_{mn} + \Phi^{(6)}_{mn} +\Phi^{(8)}_{mn} +\Phi^{(10)}_{mn} \right)
\end{equation}
(Here each $\Phi^{(2j)}_{mn}$ is a polynomial of the degree $2j$ in $dx_L$.) Indeed, as elements
of ${\cal M}$ are harmonic functions, $d_L\delta_L\Phi^{(2)}_{mn}  = 0$ implies $\delta_Ld_L \Phi^{(2)}_{mn}  = 0$ and 
therefore the existence of $\Phi^{(4)}$ such that $d_L \Phi^{(2)}_{mn} + \delta_L \Phi^{(4)}_{mn} =0 $. And so on 
until $\Phi_{mn}^{(10)}$. 

Since $\Phi^{(10)}_{mn}$ is a top form, exists $\Psi^{(9)}\in {\cal M}$ such that $\Phi^{10} = d_L\Psi^{(9)}$. 
Furthermore, $d_L(\Phi^8 - \delta_L \Psi^{(9)}) = 0$, therefore exists $\Psi^{(7)}\in {\cal M}$ such that 
$\Phi^8 - \delta_L \Psi^{(9)} = d_L\Psi^{(7)}$. Continuing, we get $\Phi^{(6)} - \delta_L \Psi^{(7)} = d_L\Psi^{(5)}$, 
$\Phi^{(4)} - \delta_L \Psi^{(5)} = d_L\Psi^{(3)}$ and finally $d_L(\Phi^{(2)} - \delta_L \Psi^{(3)}) = 0$. Let us denote:
\begin{align}
   \Phi \;& = 
\Phi^{(2)} + \Phi^{(4)} + \Phi^{(6)} +\Phi^{(8)} +\Phi^{(10)} 
\\     
\Psi \;& = \;\;\;\; \Psi^{(3)} + \Psi^{(5)} + \Psi^{(7)} + \Psi^{(9)}
\end{align}
Then we get:
\begin{equation}
     dx_L^p \;  A_{p\;;\;[m}\stackrel{\leftarrow}{\partial}_{n]} = 
(\delta_L + d_L) \Big( \Phi - (\delta_L + d_L)\Psi\Big)
\end{equation}
Notice that $\Phi - (\delta_L + d_L)\Psi$ is of the form:
\begin{equation}
   \Phi - (\delta_L + d_L)\Psi = dx_L^p\wedge dx_L^q\;\widetilde{\Psi}_{pq\;;\;mn}
\end{equation}
This concludes the proof that the ansatz (\ref{AnsatzMaxwellMaxwell}) is trivial iff $F_{[pq]\;;\;[mn]} = 0$.

\paragraph     {Case $N=0$}
The vanishing lemma (\ref{VanishingOfCohomologyDPlusDelta}) does not work in the case $N=0$, in this case the 
cohomology of $d_L$ is given by the formulas of Section \ref{sec:CohomologyInMaxwell} with the replacement
$c^m\mapsto dx^m_L$, $dx^m\mapsto dx_R^m$. Similarly, the cohomology of $\delta_L$ is obtained
via the Hodge duality. Therefore, it is necessary to repeat the analysis taking 
into account this nontrivial cohomology. 
There is no obstacle to satisfy (\ref{FromAToPhi2}), even if $A_{p\;;\;[m}\stackrel{\leftarrow}{\partial}_{n]} = \mbox{const}$,
because there are no 11-forms and therefore the cohomology of $\delta_L$ vanishes on
expressions which are monomials of the first order in $dx_L$.
There are potential obstacles in completing the chain (\ref{ChainOfPhis}). We will not do the
analysis here, but just point out that by rotational symmetry, the potential 
obstacles are proportional to the following constant tensors: the total 
antisymmetrization and the contraction:
\begin{align}
{\cal C}_{pmn} \;&= A_{[p\;;\;m}\stackrel{\leftarrow}{\partial}_{n]}
\label{TotalAntisymmC} \\  
{\cal C}_n \;&= g^{pm}A_{p\;;\;[m}\stackrel{\leftarrow}{\partial}_{n]}
\label{ContractionC}
\end{align}

\subsection{Dirac-Maxwell sector}
Consider the following ansatz for the leading term of the expasion in powers
of $x_L$:
\begin{equation}\label{DiracMaxwellAnsatz}
   \hat{c} \Psi_{[m}(x_R)\stackrel{\leftarrow}{\partial}_{n]} dx_R^m\wedge dx_R^n
\end{equation}
where $\Psi$ satisfies the Maxwell equation $\Psi\stackrel{\leftarrow}{\bf M}\;=0$. This is in the image of $d_1$ 
when exists $\Phi$ such that:
\begin{align}
   \Psi_{[m}\stackrel{\leftarrow}{\partial}_{n]} \; = \;& 
\Gamma^k\partial_k \Phi_{[m}\stackrel{\leftarrow}{\partial}_{n]}
\label{DiracMaxwellTrivialization}\\   
\mbox{\small\tt and }\;&
\Phi \stackrel{\leftarrow}{\bf M} = 0
\nonumber
\end{align}
Then it follows that $\Delta \Phi_{[\bullet}\stackrel{\leftarrow}{\partial}_{\bullet]} = 0$ and therefore:
\begin{align}
   \Gamma^m\partial_m \Psi_{[\bullet}\stackrel{\leftarrow}{\partial}_{\bullet]} = 0
\end{align}
If $\Psi$ does not satisfy this equation, then the trivialization (\ref{DiracMaxwellTrivialization}) is 
impossible. Notice that  $\Psi\stackrel{\leftarrow}{\bf M}\;=0$, therefore $\Psi_{[\bullet}\stackrel{\leftarrow}{\partial}_{\bullet]}$ is automatically 
annihilated  by $\Delta$. But it is not necessarily annihilated by the left Dirac 
operator.

We conclude that $\Gamma^m\partial_m\Psi_{[\bullet}\stackrel{\leftarrow}{\partial}_{\bullet]}$ is an obstacle for (\ref{DiracMaxwellAnsatz}) to be trivial. 

\subsection{Maxwell-Dirac sector}
Consider the following ansatz for the leading term:
\begin{align}\label{MaxwellDiracAnsatz}
   \Psi_m(x_R)\; c_ndx_L^n\wedge dx_L^m
\end{align}
where $\Psi$ satisfies the right Dirac equation $\Psi_m\stackrel{\leftarrow}{\partial}_n\Gamma^n=0$ and also 
$\partial^m\Psi_m = 0$. This is trivial if exists $A_m$ such that:
\begin{align}
   \Psi_m = \;& \partial_n(\partial_n A_m - \partial_m A_n)
\label{TrivialityOfPsiMaxwellDirac}\\   
\mbox{\tt\small with }\;& A\stackrel{\leftarrow}{\partial}_k\Gamma^k = 0
\end{align}
This implies that $\Delta A = 0$ and therefore $\partial_{[\bullet}\Psi_{\bullet]} = 0$. Therefore $\partial_{[\bullet}\Psi_{\bullet]}$ is an 
obstacle for (\ref{MaxwellDiracAnsatz}) to be trivial. For the polynomials of nonzero degree this
is the only obstacle. Indeed, suppose that $\partial_{[m}\Psi_{n]} = 0$. As the cohomology 
of Dirac solutions at the nonzero degree is zero, this implies that:
\begin{equation}
   \Psi_m = \partial_m \Xi
\end{equation}
where $\Xi = \Xi(x_R)$ satisfies the right Dirac equation. The cohomology of $H^{<9}(\delta)$
on the solutions of the Dirac equation is zero, therefore exists $A_n$ such that 
$\Xi = -\partial^n A_n$ where $\Phi_n$ satisfies the Dirac equation. This implies (\ref{TrivialityOfPsiMaxwellDirac}).

\section{Second cohomology of the tensor product of two classical electrodynamics}\label{sec:SecondCohomologyTensorProduct}
The term $E_2^{1-{M\over 2}\,,\, 1+{M\over 2}}$ is generated by two types of terms:
\begin{align}
   & c_pc_q dx_L^p\wedge dx_L^q\;A_{[m}(x_R)\stackrel{\leftarrow}{\partial}_{n]}\;
   dx_R^m\wedge dx_R^n \;+
   \nonumber \\   
   +
   & c^pc^qdx_L^r\wedge dx_L^s \;B_{[pqrs]\,;\,[m}(x_R)\stackrel{\leftarrow}{\partial}_{n]}\;
   dx_R^m\wedge dx_R^n
\end{align}
where $A_{[m}(x_R)\stackrel{\leftarrow}{\partial}_{n]}$ is a polynomial of degree $M$ in $x_R$.
Under $d_2\,:\, E_2^{-1-{M\over 2}\,,\,2+{M\over 2}}\rightarrow E_2^{1-{M\over 2}\,,\, 1+{M\over 2}}$ the first term cancels with the right hand 
side of Eq. (\ref{AnsatzMaxwellMaxwellUnobstructedA}), because $H^9({\rm Maxw})=0$.
The second term for non-constant $B_{[pqrs]\,;\,[m}\stackrel{\leftarrow}{\partial}_{n]}$ cancels with the right hand side of 
Eq. (\ref{AnsatzMaxwellMaxwellUnobstructedB}).
The constant $B_{[pqrs]\,;\,[m}\stackrel{\leftarrow}{\partial}_{n]} = \mbox{const}$ ({\it i.e.} $M=0$) generates $\Lambda^6{\bf C}^{10}$ (because
$H^4({\rm Maxw})=\Lambda^6{\bf C}^{10}$; the $d_2$ acts as $Q_{\rm Lie}$ on Maxwell solutions):
\begin{equation}
   H^2({\rm SM}_L\otimes {\rm SM}_R) = \Lambda^6{\bf C}^{10}
\end{equation}

\section{BRST cohomology}\label{sec:BRSTCohomology}
We are now ready to compute the cohomology of $Q_{\rm SUGRA}$. 
\subsection{Ghost number one}
The corresponding part of ${\cal E}_2$ consists of two parts:
\begin{align}
   {\cal E}_2^{1,0} \;&= {\bf C}^{10}
   \\     
   {\cal E}_2^{0,1} \;&= 
   H^0({\rm SMaxw}_L)\bigoplus H^0({\rm SMaxw}_R) =
   \nonumber \\  
   \;&=    \Lambda^2{\bf C}^{10} \bigoplus \Lambda^2{\bf C}^{10} 
   \bigoplus {\bf C}^{16} \bigoplus {\bf C}^{16}
\end{align}
However, there is a nontrivial $d_2\;:\; {\cal E}_2^{0,1}\;\to\; {\cal E}_2^{2,0} = \Lambda^2{\bf C}^{10}$, which cancels the 
$L\leftrightarrow R$ antisymmetric part of $\Lambda^2{\bf C}^{10} \bigoplus \Lambda^2{\bf C}^{10} \subset {\cal E}_2^{0,1}$ with ${\cal E}_2^{2,0}$. We are left 
with:
\begin{align}
   {\cal E}_{\infty}^{1,0} \;&= {\bf C}^{10}
   \\  
   {\cal E}_{\infty}^{0,1} \;&= 
   \Lambda^2{\bf C}^{10} \bigoplus {\bf C}^{16} \bigoplus {\bf C}^{16}
   \\
   {\cal E}_{\infty}^{2,0} \;&= 0
\end{align}
These vertices are in one-to-one correspondence with the generators of the
super-Poincare algebra. 

\subsection{Ghost number two}
The corresponding part of ${\cal E}_2$ consists of three parts:
\begin{align}
   {\cal E}_2^{2,0} \;&= \Lambda^2 {\bf C}^{10}
   \\        
   {\cal E}_2^{1,1} \;&= H^1({\rm SMaxw}_L) \bigoplus H^1({\rm SMaxw}_R)
   \\   
   {\cal E}_2^{0,2} \;&= H^0({\rm SMaxw}_L\otimes {\rm SMaxw}_R)
\end{align}
\subsubsection{${\cal E}_2^{2,0}$}
We have already seen that ${\cal E}_2^{2,0}$ gets killed by the $d_2$:
\begin{equation}
      {\cal E}_{\infty}^{2,0} \;= 0
\end{equation}
\subsubsection{${\cal E}_2^{1,1}$}\label{sec:calE11}
Let us look at ${\cal E}_2^{1,1}$. We have:
\begin{align}
   {\cal E}_2^{1,1} \;&= 
   \left({\bf C}^{10} \oplus \Lambda^3{\bf C}^{10} \oplus {\bf C}^{16}\right)
   \bigoplus
   \left({\bf C}^{10} \oplus \Lambda^3{\bf C}^{10} \oplus {\bf C}^{16}\right)
\end{align}
The interpretation is as follows:
\begin{itemize}
\item ${\bf C}^{10}\oplus {\bf C}^{10}$ corresponds to the linear dilaton and the 
   ``asymmetric linear dilaton'' (the nonphysical vertex of \cite{Mikhailov:2012id} with constant 
   $A_m^-$)
\item One copy of $\Lambda^3 {\bf C}^{10}$ cancels under $d_2$ with ${\cal E}_2^{3,0}$
\item Another copy of $\Lambda^3{\bf C}^{10}$ is the NSNS $B$-field strength $H=dB$
\item Two copies of ${\bf C}^{16}$ are both unphysical
\end{itemize} 

\subsubsection{${\cal E}_2^{0,2}$}\label{sec:calE02}
This was computed in Section \ref{sec:ZerothCohomology}. We identify $A_m^+$ as $\partial_m \Phi$ (the gradient of 
the dilaton) and $A_m^-$ is the unphysical state of \cite{Mikhailov:2012id}. Notice that Eq. (\ref{EqAMinus}) 
implies that $\partial_p\partial_qA_m^- = 0$, {\it i.e.} $A_m^-$ is a linear function of $x$. Notice 
that Eqs. (\ref{PhiLDef}) and (\ref{PhiRDef}) define $\phi^{L|R}$ only up to a constant, and therefore 
$A_m^{\pm}$ is defined only up to a constant. This is because linear dilaton and 
linear asymmetric dilaton have already been counted in ${\cal E}_{\infty}^{1,1}$. 

\paragraph     {Conclusion}
As expected, $F_{pq\;;\;mn}$ has the quantum numbers of the NSNS sector of the 
linearized Type IIB SUGRA, modulo some zero mode subtleties. The symmetric part 
$F_{pq\;;\;mn} + F_{mn\;;\;pq}$ corresponds to the Riemann curvature tensor $R_{[pq][mn]}$, and 
the antisymmetric part $F_{pq\;;\;mn} - F_{mn\;;\;pq}$ to $\partial_{[p} B^{\mbox{\tiny NSNS}}_{q]\;[m} \stackrel{\leftarrow}{\partial}_{n]}$. 

\subsection{Comment on nonphysical states}\label{sec:CommentOnNonPhysicalStates}
There are the following nonphysical states: 
\begin{center}
\begin{tabular}{|c|l|}
\hline
${\bf C}^{10}$ & from  ${\cal E}_{\infty}^{1,1}$: constant $A_m^-$
\\   
${\bf C}^{16}\oplus {\bf C}^{16}$ & from ${\cal E}_{\infty}^{1,1}$
\\   
$\Lambda^2{\bf C}^{10}$ & from ${\cal E}_{\infty}^{0,2}$: $\partial_{[q}A^-_{m]}$
\\  
\hline
\end{tabular}
\end{center}
They have the quantum numbers of the adjoint representation of the 
super-Poincare algebra.

In the bosonic string theory, the nonphysical states were removed by
imposing the constraint  $(b_{0} - \overline{b}_0) V = 0$ \cite{Nelson:1988ic}. This is probably possible also in
the pure spinor approach, as the pure spinor $b$-ghost was constructed in
the nonminimal formalism \cite{Berkovits:2006vi}. But there is also another way of removing the
nonphysical states, which we will now describe.

As we discussed in the Introduction, the BRST closedness of the vertex operator
is a necessary and sufficient condition for the corresponding deformation 
of the classical worldsheet action to have the classical BRST invariance.
However, at the one-loop level there is an anomaly which is cancelled by the
Fradkin-Tseytlin term \cite{Berkovits:2001ue}:
\begin{equation}\label{FradkinTseytlinTerm}
 \alpha'  \int d^2\tau\; \Phi R
\end{equation}
Here $\Phi$ is the {\em dilaton superfield}. The only place where $\Phi$ enters is the 
Fradkin-Tseytlin term (\ref{FradkinTseytlinTerm}), which does not matter at the {\em classical} level.
It is, in this sense, ``invisible'' in the classical theory. The condition of 
the one-loop BRST invariance implies that $\Phi$ is related to the ``visible'' 
superfields (those which enter in the main part of the worldsheet action) by 
some equations:
\begin{align}
   D_{\alpha}\Phi \;& = \Omega_{\alpha}
\label{DilatonLeft}\\   
D_{\hat{\alpha}}\Phi \;& = \widehat{\Omega}_{\hat{\alpha}}
\label{DilatonRight}
\end{align}
where $\Omega_{\alpha}$ and $\widehat{\Omega}_{\hat{\alpha}}$ on the right hand side are some function of the ``visible'' 
superfields. In this sense, $\Phi$ is determined, unambiguously up to a constant, 
from the ``visible'' superfields. 

However, it turns out that for some classical backgrounds the equations (\ref{DilatonLeft}) 
and (\ref{DilatonRight}) are incompatible \cite{Mikhailov:2012id}. Such backgrounds, in our terminology, are
{\em nonphysical}. Being perfectly consistent from the point of view of the
classical worldsheet sigma-model, they however fail at the one-loop level. 

\vspace{10pt}

\begin{minipage}[b]{0.45\textwidth}
This is somewhat unusual, as the typical situation is that differential 
equations are ``generally speaking incompatible, but sometimes become 
compatible''. Here we have the opposite situation. Equations (\ref{DilatonLeft}) and (\ref{DilatonRight})
for $\Phi$ are compatible for the vast majority of backgrounds, but become 
incompatible on a finite-dimensional nonphysical component. In other words, 
{\bf physical and non-physical deformations are ``mutually obstructed''}.
\end{minipage}
\begin{minipage}[b]{0.5\textwidth}
\hspace{5pt}
\includegraphics[scale=0.3]{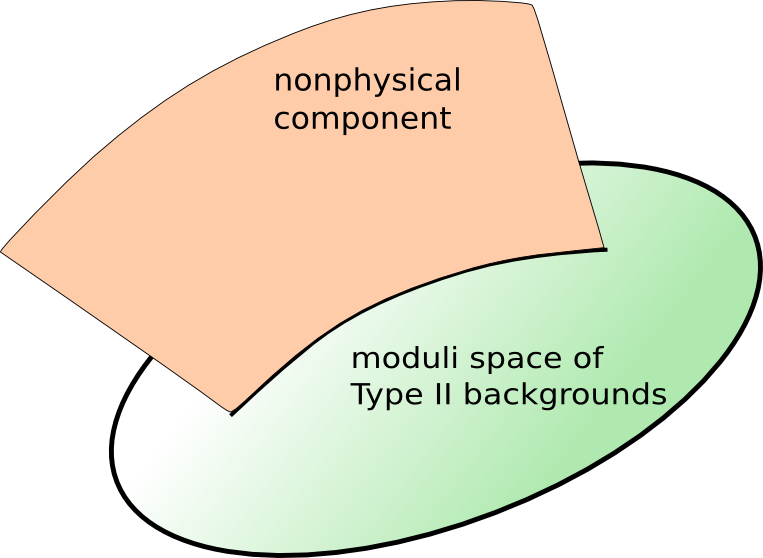}\\
\vspace{15pt}
\end{minipage}

\vspace{10pt}
\noindent 
Roughly speaking, this can be understood as follows. The compatibility
conditions for equations (\ref{DilatonLeft}) and (\ref{DilatonRight}) include the equation:
\begin{equation}
   \Gamma_m^{\alpha\beta}D_{\alpha}\Omega_{\beta} - \Gamma_m^{\hat{\alpha}\hat{\beta}}D_{\hat{\alpha}} \Omega_{\hat{\beta}} = 0
\end{equation}
Both $\Omega_{\alpha}$ and $\hat{\Omega}_{\hat{\alpha}}$ are defined in terms of other SUGRA fields, which already
satisfy the SUGRA constraints. These constraints translate into some
constraint on $\Gamma_m^{\alpha\beta}D_{\alpha}\Omega_{\beta} - \Gamma_m^{\hat{\alpha}\hat{\beta}}D_{\hat{\alpha}} \Omega_{\hat{\beta}}$ 
(which is therefore automatically satisfied).
Surprizingly, that automatic constraint seems to be 
not $\Gamma_m^{\alpha\beta}D_{\alpha}\Omega_{\beta} - \Gamma_m^{\hat{\alpha}\hat{\beta}}D_{\hat{\alpha}} \Omega_{\hat{\beta}} = 0$
but rather $\Gamma_m^{\alpha\beta}D_{\alpha}\Omega_{\beta} - \Gamma_m^{\hat{\alpha}\hat{\beta}}D_{\hat{\alpha}} \Omega_{\hat{\beta}} = \mbox{const}$, {\it i.e.} the {\em derivatives} 
of $\Gamma_m^{\alpha\beta}D_{\alpha}\Omega_{\beta} - \Gamma_m^{\hat{\alpha}\hat{\beta}}D_{\hat{\alpha}} \Omega_{\hat{\beta}}$ being zero (\cite{Mikhailov:2012uh,Mikhailov:2012id}, {\em cp.} Eqs. (\ref{ConstantFAntisymm}) and (\ref{ConstantDoubleTrace})). 
In order to kill the nonphysical component, we just have to require that this 
constant is zero; this is why the nonphysical component is finite-dimensional.

\vspace{10pt}
\noindent
We observe that the nonphysical operators seem to be in correspondence
with the global symmetries. This should have a natural interpretation in terms
of the action of the $b$-ghost:
\begin{equation}
\fbox{\mbox{nonphysical, ghost number 2}} 
\;\;\xrightarrow{b_0-\overline{b}_0}\;\;
\fbox{$\begin{array}{c}\mbox{ghost number 1}\cr 
\mbox{\small (global symmetries)}\end{array}$}
\end{equation}
But, as we explained: 
\begin{itemize}
\item instead of imposing the condition $(b_0 - \overline{b}_0)V =0$, one can request the 
   existence of the dilaton superfield $\Phi$
\end{itemize}
Notice that including $\Phi$ also solves the following problem. Our analysis, based 
on the naive BRST cohomology, failed to identify the dilaton zero mode. But 
once we include $\Phi$, the dilaton is identified as the lowest component of $\Phi$,
and in particular the zero mode of the dilaton is recovered.

\subsection{Ghost number three}
Most of the ghost number three vertex operators transform in the same 
representation as ghost number two vertex operators. This is in line with the
picture:
\begin{equation}\label{BFromThreeToTwo}
\fbox{\mbox{ghost number 3}} 
\;\;\xrightarrow{b_0-\overline{b}_0}\;\;
\fbox{\mbox{ghost number 2}}
\end{equation}
Notice that the map (\ref{BFromThreeToTwo}) lowers the polynomial degree of the vertex by 2, as
the $b$-ghost should. For example, in the Dirac-Dirac sector, the ghost number 3
vertex is of the form $\hat{c}R$; to produce the bispinor field we remove $\hat{c}$ and then
act with the left Dirac operator:
\begin{equation}
   \hat{c}R \mapsto \Gamma^m{\partial\over\partial x_R^m} R
\end{equation}
Removing $\hat{c}$ lowers the degree by one, and then ${\partial\over\partial x_R^m}$ again lowers the degree by 
one.

Let us look more carefully at the subtleties which arise when we consider 
polynomial vertices of low degree.

\subsubsection{${\cal E}_2^{3,0}$}
This is $\Lambda^3{\bf C}^{10}$. It cancels with part of ${\cal E}_2^{1,1}$ --- see Section \ref{sec:calE11}.

\subsubsection{${\cal E}_2^{2,1}$}\label{sec:E21}
This is $\Lambda^4{\bf C}^{10} \oplus {\bf C} \oplus \Lambda^4{\bf C}^{10} \oplus {\bf C}$. First of all, we have restrict to the 
kernel of $d_2\;:\; {\cal E}_2^{2,1}\to{\cal E}_2^{4,0}$. This kills one copy of $\Lambda^4{\bf C}^{10}$. But also, we have 
to take a factorspace over the image of $d_2\;:\; {\cal E}^{0,2}\to {\cal E}^{2,1}$. This  cancels 
another copy of $\Lambda^4 {\bf C}^{10}$ against the (\ref{ConstantFAntisymm}) and one copy of $\bf C$ against the  
(\ref{ConstantDoubleTrace}). For example, $d_2(dx_L^m x_L^n f_{mnpq} dx_R^p x_R^q)$ cancels the diagonal $\Lambda^4 {\bf C}^{10}$ as:
\begin{align}
   & dx_L^m x_L^n f_{mnpq} dx_R^p x_R^q \;\;\xrightarrow{Q_{\rm Lie}} 
   \nonumber\\   
   \xrightarrow{Q_{\rm Lie}}\;\; & 
   -dx_L^mc^n f_{mnpq} dx_R^p x_R^q - dx_L^mx_L^nf_{mnpq} dx_R^p c^q \;\;
   \xrightarrow{(Q_L+Q_R)^{-1}} 
   \nonumber\\   
   \xrightarrow{(Q_L+Q_R)^{-1}}\;\; &
   - x_L^mc^n f_{mnpq} dx_R^p x_R^q + dx_L^m x_L^n f_{mnpq} x_R^p c^q\;\;
   \xrightarrow{Q_{\rm Lie}} 
   \nonumber\\   
   \xrightarrow{Q_{\rm Lie}}\;\; &
   - c^mc^n f_{mnpq} dx_R^p x_R^q - dx_L^m x_L^n f_{mnpq} c^p c^q
\end{align}
and a similar computation shows that $d_2((dx_L\cdot x_R)(dx_R\cdot x_L))$ cancels a 
diagonal copy of $\bf C$. Another copy of ${\bf C}$ does not seem to cancel with anything:
\begin{equation}\label{DiscreteStateFromE21}
   {\cal E}^{2,1}_{\infty} = {\bf C}
\end{equation}

\subsubsection{${\cal E}_2^{1,2}$}
\paragraph     {Dirac-Dirac sector}
There are the following obstacles to triviality:
\begin{enumerate}
\item The bispinor $\Gamma^m{\partial\over\partial x_R^m}R$, which satisfies both left and right Dirac equations
\item There is also a discrete state (\ref{VertexGhostNumberThreeDimFour}) which corresponds to $R$ being a constant
   times the unit matrix
\end{enumerate}
\paragraph     {Maxwell-Maxwell sector}
There are the following obstacles to triviality:
\begin{enumerate}
\item The double field strength  $F_{[mn]\;;\;[pq]}$ of Section \ref{sec:DoubleFieldStrength}
\item If the double field strength is zero, then there are constant tensors
   ${\cal C}_{klm}$ and ${\cal C}_k$ defined in (\ref{TotalAntisymmC}) and (\ref{ContractionC})
\end{enumerate}
First let us look at the double field strength. Notice that the formulas of 
Section \ref{sec:DoubleFieldStrength} are almost identical to Section (\ref{sec:calE02} $\rightarrow$ \ref{sec:ZerothCohomology}). The only difference is that 
the $F_{[mn]\;;\;[pq]}$ of Section \ref{sec:ZerothCohomology} is not required to be of the form (\ref{DefineDoubleFieldStrength}) with $A_{n\,;\,p}$ 
satisfying Eq. (\ref{AnsatzMaxwellMaxwellUnobstructedA}).
As the $F_{[mn]\;;\;[pq]}$ of Section \ref{sec:DoubleFieldStrength}  {\em is} required to  be of such a form, it automatically 
satisfies:
\begin{align}
   g^{np}F_{[mn]\;;\;[pq]}(x) \;&= 
   {1\over 2} \partial_{[m} g^{np} A_{n]\;;\;[p}\stackrel{\leftarrow}{\partial}_{q]}
\end{align}
Eq. (\ref{AnsatzMaxwellMaxwellUnobstructedA}) implies the existence of $\Phi$ such that $\partial^n A_{n\;;\;p} = \partial_p\Phi$. Taking also 
into account Eq. (\ref{AnsatzMaxwellMaxwellEqm}), we get:
\begin{equation}
   g^{np}F_{[mn]\;;\;[pq]}(x) = 
   \partial_m\partial_q \left[{1\over 8}\left(g^{pn}A_{p\;;\;n} - \Phi\right)\right]
\end{equation}
This is the same equation as we got in Section (\ref{sec:ZerothCohomology}), except there is no 
unphysical $A_m^-$. 

On the other hand, there are    ${\cal C}_{klm}$ and ${\cal C}_k$ defined in (\ref{TotalAntisymmC}) and (\ref{ContractionC}), which
should be mapped by $b_0 - \overline{b}_0$ to $H^{NSNS}_{klm}$ and the dilaton gradient. Also, there 
is the discrete state (\ref{VertexGhostNumberThreeDimFour}). 

Notice that in our computation we missed the dilaton zero mode, as
the corresponding vertex is probably a BRST variation of something that is not
annihilated by $b_0 - \bar{b}_0$ \cite{Nelson:1988ic}. It is possible that the discrete state (\ref{VertexGhostNumberThreeDimFour}) mapped 
by $b_0-\bar{b}_0$ to the dilaton zero mode. However, there is also another discrete
state at the ghost number three: Eq. (\ref{DiscreteStateFromE21}). Therefore our computations seem to 
confirm Eq. (\ref{BFromThreeToTwo}), except that we see {\em two} ghost scalar ghost number three discrete
states: Eq. (\ref{VertexGhostNumberThreeDimFour}) and Eq. (\ref{DiscreteStateFromE21}).

\subsection{Ghost number four}\label{sec:GhostNumberFour}
\subsubsection{${\cal E}^{2,2}_{\infty}$}
The term ${\cal E}_2^{2,2} = H^2(Q_{\rm Lie}\;,\; {\rm SMaxw}_L\otimes {\rm SMaxw}_R)$ was computed in Section \ref{sec:SecondCohomologyTensorProduct}:
\begin{equation}
H^2({\rm SM}_L\otimes {\rm SM}_R) = \Lambda^6{\bf C}^{10}
\end{equation}
It cancels with half of:
\begin{equation}
{\cal E}^{4,1}_2=H^4({\rm SM}_L\oplus {\rm SM}_R)=\Lambda^6{\bf C}^{10}\oplus \Lambda^6{\bf C}^{10}
\end{equation}
(and another half of ${\cal E}^{4,1}_2$ then cancels with ${\cal E}^{5,0}_2 = \Lambda^6 {\bf C}^{10}$).
This pattern persists for $2<p\leq 6$, giving the short exact sequences:
\begin{align}
   0\;\longrightarrow\;& [{\cal E}^{p,2} = \Lambda^{p+4}{\bf C}^{10}]
   \;\stackrel{d_2}{\longrightarrow}\;
   \nonumber \\    
   \;\stackrel{d_2}{\longrightarrow}\;&   
   [{\cal E}^{p+2,1} =\Lambda^{p+4}{\bf C}^{10}\oplus \Lambda^{p+4}{\bf C}^{10}]
   \;\stackrel{d_2}{\longrightarrow}\;
   \nonumber \\    
   \;\stackrel{d_2}{\longrightarrow}\;&   
   [{\cal E}^{p+4,0} =\Lambda^{p+4}{\bf C}^{10}]
\end{align}

\subsubsection{${\cal E}^{4,0}_{\infty}$}
The term $H^4(Q_{\rm Lie}\;,\;{\bf C}) = \Lambda^4{\bf C}^{10}$ is nonzero, but it cancels with the $d_2$ of 
$H^2(Q_{\rm Lie}\;,\; {\rm SMaxw}_L\bigoplus {\rm SMaxw}_R)$.  

\subsubsection{${\cal E}^{3,1}_{\infty}$}
The space $\mbox{ker}
\Big( d_2:\; H^3(Q_{\rm Lie}\;,\; {\rm SMaxw}_L\bigoplus {\rm SMaxw}_R) \to H^5(Q_{\rm Lie}\;,\;{\bf C})\Big)$ is 
killed by the $d_2$ of $H^1(Q_{\rm Lie}\;,\;{\rm SMaxw}_L\otimes {\rm SMaxw}_R)$. Indeed, let us consider 
the following element of $H^1(Q_{\rm Lie}\;,\;{\rm SMaxw}_L\otimes {\rm SMaxw}_R)$ with constant 
$B_{pqr\;;\;[m}\stackrel{\leftarrow}{\partial}_{n]}$:
\begin{equation}
   c^pdx_L^q\wedge dx_L^r\; B_{pqr\;;\;[m}\stackrel{\leftarrow}{\partial}_{n]}\; dx_R^m\wedge dx_R^n
\end{equation}
(This is a particular case of (\ref{AnsatzMaxwellMaxwell}) with zero $A$ and constant $B$.) Being an 
element of  $H^1(Q_{\rm Lie}\;,\;{\rm SMaxw}_L\otimes {\rm SMaxw}_R)$, this is a $c$-dependent element
of the cohomology of $Q_L + Q_R$, parametrized by a left times right field 
strength. We need to act on this by the $d_2\;:\;{\cal E}^{1,2}_2 \to {\cal E}^{3,1}_2$. For that, we need to 
know the actual ($c$-dependent) vertex, which is built using the left and right 
{\em vector potentials}, {\it i.e.} $c^p\;x_L^q (\theta_L\Gamma^r\lambda_L)\;B_{pqr\;;\;[m}\stackrel{\leftarrow}{\partial}_{n]}\; x_R^m (\theta_R\Gamma^n\lambda_R)$. 
The $Q_{\rm Lie}$ on the vertex is not zero:
\begin{equation}
   - c^pc^q (\theta_L\Gamma^r\lambda_L) B_{pqr\;;\;[m}\stackrel{\leftarrow}{\partial}_{n]} x_R^m(\theta_R\Gamma^n\lambda_R) - c^p x_L^q (\theta_L\Gamma^r\lambda_L) B_{pqr\;;\;[m}\stackrel{\leftarrow}{\partial}_{n]} c^m(\theta_R\Gamma^n\lambda_R)
\end{equation}
but is a pure gauge, namely it is $Q_L + Q_R$ of:
\begin{equation}
-  c^p c^q x_L^r B_{pqr\;;\;[m}\stackrel{\leftarrow}{\partial}_{n]} x_R^m (\theta_R\Gamma^n\lambda_R) 
+ c^p x_L^q (\theta_L\Gamma^r\lambda_R) B_{pqr\;;\;[m}\stackrel{\leftarrow}{\partial}_{n]} c^m x_R^n
\end{equation}
And the $Q_{\rm Lie}$ of this gives:
\begin{align}
- c^pc^qc^r B_{pqr\;;\;[m}\stackrel{\leftarrow}{\partial}_{n]}x_R^m(\theta_R\Gamma^n\lambda_R) 
+ c^p x_L^q (\theta_L\Gamma^r\lambda_R) B_{pqr\;;\;[m}\stackrel{\leftarrow}{\partial}_{n]} c^m c^n +
\\    
+ c^pc^q x_L^r B_{pqr\;;\;[m}\stackrel{\leftarrow}{\partial}_{n]} c^m (\theta_R\Gamma^n\lambda_R) 
- c^p c^q (\theta_L\Gamma^r\lambda_L) B_{pqr\;;\;[m}\stackrel{\leftarrow}{\partial}_{n]} c^m x_R^n
\end{align}
The second row is $-(Q_L+Q_R)c^pc^q x_L^r B_{pqr\;;\;mn} c^m x_R^n$. And the first row is 
equivalent, in the Maxwell cohomology, to the expression:
\begin{equation}
   -c^pc^qc^r{\cal B}_{[pqrmn]} (dx_R^m\wedge dx_R^n - dx_L^m\wedge dx_L^n)
\end{equation}
where  ${\cal B}_{[pqrmn]}$ is defined in (\ref{DefBpqrmn}).
This can be used to kill any class of the form:
\begin{equation}
   c^pc^qc^r G_{[pqrst]}(dx_R^s\wedge dx_R^t - dx_L^s\wedge dx_L^t)
\end{equation}
in $H^3(Q_{\rm Lie}\;,\; {\rm SMaxw}_L\bigoplus {\rm SMaxw}_R)$. The classes of the form:
\begin{equation}
   c^pc^qc^r H_{[pqrst]}(dx_R^s\wedge dx_R^t + dx_L^s\wedge dx_L^t)
\end{equation}
are not in the image of $d_2$. However, the $d_2$ of them is nonzero, giving
an element of ${\cal E}^{5,0}_2 = H^5({\bf C}^{10})$ of the form $c^pc^qc^rc^sc^t H_{pqrst}$.

We conclude that ${\cal E}_3^{3,1} = 0$.

\section{Action of the supersymmetry on the ghost number three vertices}\label{sec:ActionOfSUSY}
In this section we will study the action of the supersymmetry on the ghost
number three vertices. We will first act by the left supersymmetry on the
element of the Maxwell-Dirac sector, an see that the result is some element of
the Dirac-Dirac sector. Then we will act my the left supersymmetry on the
Dirac-Dirac sector, which will bring us back to the Maxwell-Dirac sector. We 
will verify that the anticommutator of two supersymmetries is a translation.

\subsection{Left supersymmetry on the Maxwell-Dirac sector}
Let us consider an element of the Maxwell-Dirac sector, of the following form:
\begin{align}
   \Psi_m(x_R)c_ndx_L^n\wedge dx_L^m + \ldots
\end{align}
where $\ldots$ stand for elements of the lower degree in $x_R$ (which have dependence 
on $x_L$). Let us act on it by the {\em left} supersymmetry with the parameter 
$\epsilon^{\alpha}$, which we will call $S_{\epsilon}$. To evaluate the action of this supersymmetry, we 
will use the formulas from Section 6.1.3 of \cite{Mikhailov:2009rx} (where $S_{\epsilon}$ was denoted $Q_{\rm Lie}^{\cal H}$, 
and $\epsilon^{\alpha}$ was called $\xi^{\alpha}$). We get the following element of the Dirac-Dirac sector:
\begin{equation}\label{SUSYOnPsiGivesPreR}
 -{2\over 3}\times  {1\over 2} [\hat{c},\Gamma^m]\epsilon\;\Psi_m + \ldots
\end{equation}
We observe:
\begin{equation}
   \Gamma^j{\partial\over\partial x_L^j} \left(   
      {1\over 2} [\hat{x}_L,\Gamma^m]\epsilon\;\Psi_m 
   \right) = \Gamma^j{\partial\over\partial x_L^j}\left(
      {9\over 10}    \hat{x}_L\Gamma^m\epsilon\Psi_m
   \right)
\end{equation}
This implies that (\ref{SUSYOnPsiGivesPreR}) gives the same cohomology class as:
\begin{equation}
-{2\over 3}\times{9\over 10}  \hat{c}\Gamma^n\epsilon\; \Psi_n(x_R) + \ldots
\end{equation}
In notations of Section \ref{sec:DefineR} we have: 
\begin{equation}\label{SUSYOnPsiGivesR}
R= -{2\over 3}\times{9\over 10}\Gamma^n\epsilon\;\Psi_n(x_R) 
\end{equation}
The obstacle to the triviality is:
\begin{equation}
   \Gamma^m\partial_m R  = 
 -{2\over 3} \times {9\over 10}\Gamma^m\Gamma^n\epsilon\; \partial_m\Psi_n(x_R)
= -{2\over 3} \times{9\over 10}\Gamma^{mn}\epsilon\; \partial_{[m}\Psi_{n]}(x_R)
\end{equation}
(This is a bispinor: $(\Gamma^{mn}\epsilon)^{\alpha}\; (\partial_{[m}\Psi_{n]}(x_R))^{\dot{\beta}}$)

\subsection{Left supersymmetry on the Dirac-Dirac sector}
We want to calculate the action of the supersymmetry with the parameter $\epsilon$ on 
the class:
\begin{equation}
   \hat{c}R (x_R) + \ldots
\end{equation}
This is a bit tricky, becase the leading term $\hat{c}R(x_R)$ does not contribute,
and we have to analyze the subleading term proportional to $x_L$:
\begin{align}
 \;&  \hat{c}R(x_R) \;+
\nonumber\\   
+ \;& 
{5\over 6}\hat{c} x_L^k\partial_kR(x_R) - 
{1\over 6} \hat{x}_L c^k\partial_k R(x_R) + 
{5\over 6} (c\cdot x_L)\Gamma^k\partial_k R \;+ \ldots
\end{align}
where $\ldots$ stand for terms of the higher order in $x_L$. Again, we use the formulas
from  \cite{Mikhailov:2009rx}. When acting by the supersymmetry with the parameter $\epsilon$, we are 
getting the following element of the Maxwell-Dirac sector:
\begin{align}
-{3\over 2}   \epsilon \Gamma_{[n}\left( 
      {5\over 6}\hat{c}\partial_{m]}R(x_R) -
      {1\over 6}\Gamma_{m]}c^k\partial_kR(x_R) + 
      {5\over 6}c_{m]}\Gamma^k\partial_kR(x_R)
   \right)dx_L^n\wedge dx_L^m
\end{align}
This can be written as follows:
\begin{align}
&   -{3\over 2}\; c^kY_{kmn}^l\partial_lR\;dx_L^n\wedge dx_L^m
\nonumber\\   
\mbox{\tt\small where } \;&  Y_{kmn}^l = 
- {5\over 6} \Gamma^l{}_{[n}\delta_{m]k} 
- {5\over 6} \Gamma_{k[n}\delta^l_{m]}
- {1\over 6} \Gamma_{nm}\delta^l_k
\end{align}
We can add $Q_{\rm Lie}\left(\epsilon \Gamma_{nm}R dx_L^n\wedge dx_L^m\right)$ then we get:
\begin{align}
-{3\over2}\;   c^k\widetilde{Y}_{kmn}^l\partial_lR\;dx_L^n\wedge dx_L^m
\nonumber\\   
\mbox{\tt\small where } \widetilde{Y}_{kmn}^l = 
- {5\over 6} \Gamma^l{}_{[n}\delta_{m]k} 
- {5\over 6} \Gamma_{k[n}\delta^l_{m]}
+ {5\over 6} \Gamma_{nm}\delta^l_k
\end{align}
Now $\widetilde{Y}^l_{[kmn]}=0$ and $\widetilde{Y}^l_{mmn} = -5 \Gamma^l{}_n$. Consider the following tensor:
\begin{equation}
   Z^l_{kmn} = {5\over 18} \Gamma^l{}_{[n}\delta_{m]k}  
- {5\over 6} \Gamma_{k[n}\delta^l_{m]}
+ {5\over 6} \Gamma_{nm}\delta^l_k
\end{equation}
It satisfies  $Z^l_{[kmn]}=0$ and  $Z^l_{mmn}=0$. Therefore the cohomology class
does not change if we replace $\widetilde{Y}^l_{kmn}$ as follows:
\begin{equation}
   \widetilde{Y}^l_{kmn} \mapsto \widetilde{Y}^l_{kmn} - Z^l_{kmn} = -{10\over 9} \Gamma^l{}_{[n}\delta_{m]k}  
\end{equation}
Indeed: 
\begin{align}
\;& c^k Z_{kmn}^l\partial_lR\;dx_L^n\wedge dx_L^m + \ldots\;= 
\nonumber \\   
=\;&
c^m\left({\partial\over\partial x_L^m} - {\partial\over\partial x_R^m}\right) 
\left(
x_L^kZ_{kmn}^l\partial_lR\;dx_L^n\wedge dx_L^m + \ldots
\right)
\end{align}
We conclude that the supersymmetry with the parameter $\epsilon$ brings $\hat{c}R+\ldots$
to $-{10\over 9} \times \left(-{3\over 2}\right)\epsilon \Gamma_{ln}\partial^l R \;c_kdx^k\wedge dx^n$. When $R$ is given by (\ref{SUSYOnPsiGivesR}), we get:
\begin{align}
   - \epsilon \Gamma_{ln} \Gamma^j\epsilon\partial_R^l\Psi_j c_kdx^k\wedge dx^n = - (\epsilon \Gamma^l \epsilon)\; {\partial\over\partial x_R^l}
(\Psi_j(x_R) c_kdx_L^k\wedge dx_L^j)
\end{align}
This is in agreement with the fact that the anticommutator of two SUSY 
transformations is a translation.

\subsection{Conclusion}
Ghost number three vertices transform in the linearized Type IIB supergravity
supermultiplet.

\section*{Acknowledgments}
We would like to thank Nathan Berkovits for useful discussions.
This work was supported in part by the Ministry of Education and Science of 
the Russian Federation under the project 14.740.11.0347 ``Integrable and 
algebro-geometric structures in string theory and quantum field theory'', and 
in part by the RFBR grant 15-01-99504 ``String theory and integrable systems''.


\def\cprime{$'$} \def\cprime{$'$}
\providecommand{\href}[2]{#2}\begingroup\raggedright\endgroup

\end{document}